\documentclass[aps,pre,preprint,groupedaddress,
,superscriptaddress,showpacs,a4paper,floatfix]{revtex4-1}

\usepackage{amsmath,amssymb,graphicx,graphics,bm}
\usepackage{hyperref}
\usepackage{graphics,graphicx}
\usepackage{amsmath,amssymb}
\newcommand{\mi}{\mathrm{i}}

\begin{document}

\title{Fluid-particle interactions and fluctuation-dissipation relations III -
Correlated fluctuations, regularity and added mass}

\author{Massimiliano Giona}
\email{massimiliano.giona@uniroma1.it}
\affiliation{Dipartimento di Ingegneria Chimica Materiali Ambiente\\
Facolt\`{a} di Ingegneria Industriale, La Sapienza Universit\`{a} di Roma,
via Eudossiana 18, 00184, Roma, Italy}

\author{Giuseppe Procopio}
\email{giuseppe.procopio@uniroma1.it}
\affiliation{Dipartimento di Ingegneria Chimica Materiali Ambiente\\
Facolt\`{a} di Ingegneria Industriale, La Sapienza Universit\`{a} di Roma,
via Eudossiana 18, 00184, Roma, Italy}

 \author{Chiara Pezzotti}
\email{chiara.pezzotti@uniroma1.it}
\affiliation{Dipartimento di Ingegneria Chimica Materiali Ambiente\\
Facolt\`{a} di Ingegneria Industriale, La Sapienza Universit\`{a} di Roma,
via Eudossiana 18, 00184, Roma, Italy}

\date{\today}

\begin{abstract}
The fluctuation-dissipation theory is grounded on the Langevin condition
expressing the  local independence between  the thermal force and the particle velocity history.
Upon hydrodynamic grounds, it is reasonable to relax this condition in
order to account for the correlated fluid fluctuations, especially in
the case of  liquids, consistently with the inclusion of acoustic effects and with
the finite speed of propagation of internal shear stresses. We show that 
the introduction of
correlated stochastic processes in the basic fluctuational
patterns defined in Giona et al. (2024), preserves the global fluctuation-dissipation
relation, connecting diffusivity to the global friction
factor, and the resulting  velocity fluctuations become almost
everywhere smooth functions of time. Moreover, a fluctuational added mass
arises as a consequence of correlations.
This leads to a fluctuation-inertia relation, connecting
the fluctuational added mass at microscale to its occurrence
for macroscopic objects.
\end{abstract}

\maketitle

\section{Introduction}
\label{sec1}

In \cite{part1} and \cite{part2}, henceforth referred to
as part I and part II, respectively, we have followed the
classical Kubo approach to fluctuation-dissipation relations, 
assuming that
the thermal force ${\bf R}(t)$ at time $t$ is uncorrelated to particle
velocity at previous time instants,
\begin{equation}
\langle {\bf R}(t) \otimes {\bf v}(0) \rangle =0 \, , \qquad t \geq 0
\label{eq1}
\end{equation}
where $\otimes$ indicates the tensor product.
This is indeed the Langevin condition \cite{langevin,kubo2},
that  undoubtedly displays analytical advantages, as it provides
straightforwardly the evolution equation for the
velocity autocorrelation function, and suggests  the 
white-noise nature of the stochastic forcings entering the
various fluctuation-dissipation patterns introduced in part I,
 typically regarded as distributional derivatives of independent
Wiener processes.

The Langevin condition eq. (\ref{eq1}) and the equipartition
relation for the squared  variance of the velocity entries
at equilibrium
\begin{equation}
\langle v_i^2 \rangle = \frac{k_B \, T}{m} \, , \qquad i=1,2,3
\label{eq2}
\end{equation}
where $m$ is the particle mass and $T$ the constant temperature of the
fluid medium (coinciding with that of the particle), are the two
foundative properties deriving from statistical physics,
defining particle velocity fluctuations and characterizing
the  Kubo fluctuation-dissipation theory.

In recent years, stort-time and high-resolution experiments on
Brownian particles both in liquids and gases \cite{exp1,exp2,exp3,exp4}
have qualitatively demostrated
the correctness of 
the scaling of the velocity autocorrelation function with time
predicted by the  Kubo theory, supplementing it  with the
hydromechanic analysis of the force exterted by the fluid onto the
particle including also the fluid-inertial effects, i.e. the Basset
force \cite{hydro1,hydro2,hydro3,hydro4}. On the other hand, these experiments have shown that Brownian
motion trajectories and velocities at short time scales in Newtonian fluids
(water, acetone) are smooth functions of time and more regular
than what expected at the Einstein's times \cite{exp4,deemed}.  Moreover,
deviations from eq. (\ref{eq2}), i.e. from the equipartition result, have been
reported \cite{optics_express}.

The latter phenomenon can be attributed hydrodynamically to the 
added mass effect, that occurs as an instantaneous fluid-inertial
contribution in incompressible fluids, proportional to the  actual
value of the particle acceleration \cite{landau,kim}.
However, as correctly observed by Zwanzig \cite{zwanzig1}, this
instantaneous effect could be  the consequence of the simplified
and unphysical assumption of the infinite sound velocity subsumed
in the assumption of incompressibility. The analysis performed
by Chow and Hermans \cite{chowhermans}, considering a compressible 
time-dependent Stokes equation for the fluid (i.e., a linear
proportionality between pressure and density) supports this thesis.
Therefore, the possible occurrence of a non-vanishing added-mass
influencing the squared velocity  variance, i.e.,
\begin{equation}
\langle v_i^2 \rangle = \frac{k_B \, T}{m+ m_a} \, , \qquad i=1,2,3
\label{eq3}
\end{equation}
with $m_a>0$, instead of eq.  (\ref{eq2}), has become a crucial 
issue both theoretically and experimentally \cite{jarz}, with
relavant implications in the modeling of hydrodynamics at microscale
\cite{giona_ns},
and in the foundations of  statistical physics.

The inclusion of the ``tiny acoustic effects'' (i.e. compressibility)
to explain the statistical properties of micrometric particles in
still fluids (Brownian motion) clearly indicates that the removal of
the unphysical assumption of infinite sound velocity  determines
significant changes in the qualitative statistical properties of the
particle velocity fluctuations. In turn, it  
necessarily implies the occurrence of 
correlated  pressure fluctuations
possessing a bounded velocity of propagation.
 It is important to underline
that the results by Chow and Hermans are grounded on the
classical approach  to fluctuation-dissipation relations,
namely on the validity of eq. (\ref{eq1})
and on the representation of the stochastic forcings as white-noise
processes.

Assuming, as physically correct and statistically relevant,
this  ''regularity ansatz'' of the acoustic effects,  an equivalent regularity
ansatz should be extended, by consistency,  to the other contributions involved in
the fluctuation-dissipation characterization of particle
dynamics. Specifically, the  simplification of $\delta$-correlated
stochastic forcing (white-noise processes), 
should  be removed on equal footing. And
this, ultimately questions the physical validity  of the Langevin condition eq. (\ref{eq1}), as an
acceptable approximation at least for liquids.

The inclusion  of correlated and more regular stochastic forcings, replacing
the $\delta$-correlated distributional derivatives of Wiener processes, 
in particle hydromechanic equations is also
consistent with the hydrodynamic picture of fluid-inertial effects,
giving rise to coherent and correlated motion
of fluid elements in the neighbourhood of the moving solid
object (the Darwin paths) \cite{darwin}.
It is also consistent with a principle of mutual coupling between
physical systems: the thermal force ${\bf R}(t)$ originates  also
from the velocity fluctuations occurring in the fluid medium
 due to the  motion of the colloidal  particle, and therefore it is
expected,  especially in liquids, that ${\bf R}(t)$ should be
correlated with the previous  velocity history of the particle itself.
In this regard, the Langevin condition eq. (\ref{eq1}) should
be viewed as a long-term approximation for time-scales much larger than
the characteristic dissipation and internal stress propagation times.

However, the inclusion of correlated forcings in the particle
equation of motion and ultimately in fluctuation-dissipation
theory, should be performed with great caution, and should be compatible with
some fundamental form of fluctuation-dissipation condition confirmed
by the experiments. This condition, at equilibrium, is given by
the global fluctuation-dissipation relation, that enabled Einstein
to  extract from the analysis of Brownian motion trajectories, the Avogadro
number, subsequently  confirmed experimentally by Perrin and by
many others \cite{einstein,perrin,expperrin}.

In this article we show that, introducing correlated stochastic
forcings within the basic fluctuational patterns defined
in part I, the  global fluctuation-dissipation relation is 
preserved under very loose assumptions (essentially related to a normalization
condition) in any
complex fluid medium for which particle diffusion is asymptotically
regular (i.e., the particle mean square displacement 
is asymptotically a linear
function of time).
The effect of correlated stochastic forcings within the fluctuational patterns is
twofold: (i) compared to the white-noise driven counterparts, a transition
is observed from fractal to regular velocity fluctuations; (ii)
correlations induce the occurrence of an added mass in any fluid dynamic
conditions (including the case of compressible hydrodynamics).
These results are indeed in agreement with the recent experimental
results on highly resolved Brownian motion in liquids \cite{exp4,deemed,optics_express}.

Within the  extension  of fluctuation-dissipation theory  developed
in this article, the
added mass is a consequence of the correlated hydrodynamic fluctuations
in a macroscopically still fluid phase, induced by the  high-frequency
erratic motion of micrometric particles.
A delicate issue arises in providing the connection between the fluctuation-induced added mass at microscale (i.e. for micrometric particles) and the
``instantaneous'' added mass observed for macroscopic objects 
(from millimetric particles to submarines and transoceanic ships), as consequence
of fluid inertia \cite{added1,added2,added3}. This connection is addressed  in Section \ref{sec6},  providing
a new form of constraint referred to as {\em the fluctuation-inertia
relation}.

The article is organized as follows. Section \ref{sec2} introduces the
hydrodynamics of the problems and the importance of removing
all the approximations associated with infinite propagation velocity
of hydrodynamic fields in the analysis of microparticle
velocity fluctuations.
Section \ref{sec3} addresses the inclusion of correlated stochastic
forcings within the basic fluctuational patterns  introduced in part I,
and derives for this class of models the validity of the
global fluctuation-dissipation relation followed, as an aftermath, by the occurrence of a fluctuational
added mass.
Section \ref{sec4}  discusses some prototypical examples.
The regularity of velocity fluctuations emerging from the 
correlated fluctuational patterns is the main focus of
Section \ref{sec5}. More precisely,
the presence of fluid inertia necessarily induces in the
classical fluctuation-dissipation paradigm (based on
white-noise stochastic forcings) an almost-everywhere singular
(fractal) structure of velocity fluctuations that becomes regularized 
whenever correlated forcings are accounted for
in these patterns.
Section \ref{sec6} addresses the connection between the fluctuational 
added mass deriving from the present theory and the instataneous added
mass of incompressible hydrodynamics referred to macroscopic objects
accelerating 
in a fluid. The concluding Section addresses some open issues and the
experimental scrutiny of the theory.

\section{Statement of the problem: hydrodynamic considerations}
\label{sec2}

The  advances in the theory of Brownian motion,
as the paradigmatic problem involving the interaction between hydrodynamics
and thermal fluctuations leading to the formulation of the 
fluctuation-dissipation theory, have  been marked by a progressive 
regularization
of the hydrodynamic description of the fluid medium, removing
the paradoxes associated with the infinite propagation speed 
of the hydrodynamic fields.

Owing to linearity, particle hydromechanics in the free space
(i.e. in the absence of  constraints deriving  from confinement,
see part I) can be always expressed as 
\begin{equation}
m \, \frac{d {\bf v}(t)}{d t}= {\bf F}_{f \rightarrow p}[{\bf v}(t)]
+ {\bf R}(t)
\label{eqxx1}
\end{equation}
where ${\bf F}_{f \rightarrow p}[{\bf v}(t)]$ is the mean field 
force exerted by
the fluid onto the particle, deriving from the hydrodynamic equations, and
${\bf R}(t)$ the thermal force possessing zero mean,
$\langle {\bf R}(t) \rangle =0$.
The term  ${\bf F}_{f \rightarrow p}[{\bf v}(t)]$  is a
linear functional of the history of particle velocity  from the initial time to
the actual time $t$.

These advances  towards an increasingly accurate hydrodynamic
description of the fluid rheology and of the fluid-particle
interactions correspond to the complexification
of the fluctuational patterns discussed in part I,
say from ${\mathbb S}_{m,E}^{(1,1,0)}$, where $E=k_B \, T$,
to ${\mathbb D} {\mathbb I}_{m,E}^{(3,N_d,N_i)}$, i.e.
a three-dimensional tensor model including dissipative
and fluid-inertial memory effects via $N_d$ and $N_i$ modes, respectively, 
 in which either the complex geometry of the object,
the generic rheology of the fluid medium, and the time-dependent hydrodynamics
of the fluid velocity field are taken into account.

The Einstein-Langevin picture refers to an incompressible Stokes
hydrodynamics in which the reaction of the fluid to the 
perturbation associated with the motion of a solid object immersed in it
is instantaneous (infinite speed of propagation of velocity perturbations).
Thus, for a spherical particle of radius $R_p$, the force exerted by the fluid on the particle, 
\begin{equation}
{\bf F}_{f \rightarrow p}[{\bf v}(t)] = - 6 \, \pi \, \mu \, R_p \, {\bf v}(t)
\label{eqxx2}
\end{equation}
is a function of the actual velocity value ${\bf v}(t)$, where
$\mu$ is the fluid viscosity.
  
The hydrodynamic theory of Brownian motion developed in the '70 of the
last century  focused and improved the paradoxes (in the meaning by 
Birkhoff \cite{birkhoff}) of this representation. The instantaneous
Stokes description was replaced by  the time-dependent incompressible
Stokes hydrodynamics. This leads for a Newtonian  fluid to the
following expression for the force ${\bf F}_{f \rightarrow p}[{\bf v}(t)]$
at time $t$
\begin{eqnarray}
{\bf F}_{f \rightarrow p}[{\bf v}(t)]
&= & - 6 \, \pi \, \mu \, R_p \, {\bf v}(t) \underbrace{- 6 \sqrt{\pi \, \mu
\, \rho_0} R_p^2 \int_0^t \frac{1}{\sqrt{t-\tau}} \left (
\frac{d {\bf v}(\tau)}{d \tau} + {\bf v}(0) \, \delta(\tau) \right ) \, d \tau}_{Basset \; force} \nonumber \\
&\,& \underbrace{ - \frac{2}{3} \pi \, R_p^3 \, \rho_0 \, \frac{d {\bf v}(t)}{d t}}_{added-mass \; term}
\label{eq2_1}
\end{eqnarray}
where  $\rho_0$ is the density of the fluid,  in which the Basset force emerges as a memory integral over the particle velocity history and the instantaneous added-mass term arises, proportional
to the actual value of the particle acceleration.
Expressed via a memory kernel, the added-mass term corresponds to an
impulsive  fluid-inertial 
kernel, proportional to a Dirac $\delta$-distribution centered
at $t=0$.

Also the instantaneous added-mass term is a consequence of the
paradox of infinite propagation velocity of the pressure field, intrinsic to the assumption of incompressibility. Incompressibility overlooks acoustic effects and corresponds to an infinite speed
 of sound in the fluid \cite{landau}.

The removal of the latter infinite-velocity paradox (and consequently
the inclusion of acoustics within the particle hydromechanics) has been considered by
Chow and Hermans \cite{chowhermans}, expressing   compressibility
effects and finite sound speed in the form
of a linear relation between pressure $P$ and density $\rho$  (of the
fluid), that in the present case is no longer constant,
\begin{equation}
\frac{d P}{d \rho}= c_s^2
\label{eq2_2}
\end{equation}
where $c_s$ is the sound velocity, considering
the linearized hydrodynamics for the fluid velocity field ${\bf u}({\bf x},t)$,
\begin{eqnarray}
\rho_0 \, \frac{\partial {\bf u}}{\partial t} & = & \mu \, \nabla^2 {\bf u} 
+ \left ( \frac{\mu}{3} + \kappa \right ) \nabla \left (\nabla \cdot {\bf u}
\right ) - \nabla P \nonumber \\
\frac{\partial \rho}{\partial t} & = & - \rho_0 \, \nabla \cdot {\bf u}
\label{eq2_3}
\end{eqnarray}
where $\rho_0$ is the constant density of the fluid at rest, and $\kappa$ the bulk (compression) viscosity.

Solving the hydrodynamic problem in the free space equipped
with no-slip conditions at the external surface $S_p$ of the
spherical particle, ${\bf u}({\bf x},t)|_{{\bf x} \in S_p}={\bf v}(t)$,
and evaluating the force exerted by the fluid on the particle, the added-mass
effect disappears as an instantaneous contribution. More precisely, the hydromechanics induced
by fluid-inertial effects is described by a convolution integral 
with respect to the particle acceleration defined by a smooth continuous
kernel $k(t)$ not containing any impulsive (Dirac-$\delta$) contribution.

The hydrodynamic pressure $P$ accounts for compressible internal
stresses in the fluid, that superimpose to the shear stresses.
Including finite sound velocity and compressibility implies, by
consistency, that also the   paradox of infinite propagation velocity
should be removed from the shear stress   dynamics,  which resides
in the Newtonian constitutive equation of direct (and instantaneous)
proportionality between the shear-stress tensor and the deformation
tensor. 
The simplest way for achieving this goal is to consider
a linear viscoelastic constitutive equation characterized
by a single relaxation rate (Maxwell fluid) \cite{franoschmaxwell}.

In point of fact, the Newtonian constitutive equations are an
approximation corresponding to the instantaneous response of the
internal shear stresses to  deformations. Even
water at room temperature admits a viscoelastic behavior
characterized  by a relaxation rate order of 1 ps \cite{ruocco}.

Performing this extension (in the incompressible case) \cite{pgfluids2},
the
Basset inertial contribution becomes a memory
integral,
\begin{equation}
{\bf F}_{f \rightarrow p}[{\bf v}(t)] |_{\rm Basset}= - \int_0^t k(t-\tau)
\, \left (\frac{d {\bf v}(\tau)}{d \tau} + {\bf v}(0) \, \delta(\tau) \right )
\, d \tau 
\label{eq2_4}
\end{equation}
where
\begin{equation}
k(t)=\frac{6 \, \sqrt{\pi \, \mu \, \rho_0} \, R_p^2}{\sqrt{\tau_c}} \, e^{-t/2 \tau_c} \, I_0 \left (
\frac{t}{2 \, \tau_c} \right )
\label{eq2_4x}
\end{equation}
where $I_0(\xi)$ is the modified Bessel functionof the first kind of zeroth order.
and the singularity at $t=0$ disappears.
The regularization of the Basset kernel  via the  inclusion
of viscoelastic constitutive equations, is the typical manifestation of
the removal of the paradoxes and infinities associated with the 
infinite speed of propagation  that characterizes all the parabolic
transport schemes \cite{jou,giona_pucci}.

Starting from this hydrodynamic premise, it is  consequential
 to extend the
regularity issue 
to all the contribution entering the fluctuation-dissipation patterns,
removing in this way
the paradoxes of infinite propagation velocity and their unphysical
consequences.
According to this program, it is  more
realistic, as supported by the above mentioned 
hydrodynamic observations, to formulate
 a fluctuation-dissipation theory in which the
stochastic forcing terms entering the fluctuational
patterns would possess some colored characters (i.e. continuous
and almost everywhere smooth correlation functions),
 instead of white-noise
processes, such as the distributional derivatives of Wiener processes.
As a consequence of this extension, the thermal force ${\bf R}(t)$ may
be  correlated with ${\bf v}(\tau)$ for $\tau \leq t$, i.e.
\begin{equation}
\langle {\bf R}(t) \otimes {\bf v}(0) \rangle \neq 0 \, ,  \qquad t \geq 0
\label{eq2_6}
\end{equation}
corresponding to the breaking of the Langevin condition eq. (\ref{eq1}).
Again, the hydrodynamic phenomenology supports this generalization:
inertial
effects and  the finite sound velocity of acoustics in a fluid medium,
push for  the adoption of a corresponding property (of bounded propagation speed) as regards
the stochastic forcings acting either on
${\bf v}(t)$ or on the auxiliary variables characterizing the local representation of the memory dynamics.
However, the choice of the stochastic forcing ${\bf R}(t)$ could  not  be arbitrary,
 as it should fulfill some basic properties characterizing thermal equilibrium,
expressed by the global fluctuation-dissipation relation. 
This can be achieved by
resorting to the basic fluctuational patterns introduced in part I, 
substituting the white-noise processes with correlated ones.

\section{Correlated fluctuations in basic patterns: main properties}
\label{sec3}

In  extending fluctuation-dissipation theory
beyond the Langevin condition, the founding principle is the
validity of the Einsteinian global fluctuation-dissipation relation
for a spherical particle
\begin{equation}
D \, \eta = k_B \, T
\label{eq3_1}
\end{equation}
relating the particle diffusion coefficient $D$ to the
Stokes friction factor $\eta$ in a Newtonian fluid at constant temperature
$T$.

Consider the case of a generic fluid characterized by
linear rheological constitutive equation and a particle
of arbitrary shape.
The force acting on the particle is expressed by  
\begin{equation}
{\bf F}_{f \rightarrow p}[{\bf v}(t)] = - {\bf h}(t) *
{\bf v}(t) - {\bf k}(t) * \left ( \frac{d {\bf v}(t)}{d t}+ {\bf v}(0) \, \delta(t) \right )
\label{eq3_2}
\end{equation}
where in general the dissipative ${\bf h}(t)$ and
the fluid-inertial ${\bf k}(t)$ kernels have a tensorial character.
As a byproduct, also particle motion becomes anisotropic and characterized
by a diffusivity tensor ${\bf D}$. Henceforth ``$*$''  will indicate convolution.

The fluid inertial kernel does not contribute to dissipation, surely in the case of regularly diffusive case,
and the extension of the Stokes-Einstein relation for the
particle diffusion tensor ${\bf D}$
becomes
\begin{equation}
{\bf D}  \, \boldsymbol{\eta}_\infty = k_B \, T  \, {\bf I}
\label{eq3_3}
\end{equation}
where ${\bf I}$ is the $3 \times 3$ identity
matrix and the long-term  friction factor   $\boldsymbol{\eta}_\infty$
is expressed by
\begin{equation}
\boldsymbol{\eta}_\infty = \int_0^\infty {\bf h}(t) \, d t
\label{eq3_4}
\end{equation}
If $\boldsymbol{\eta}_\infty $ is bounded, particle diffusion in the
long-term is regular (i.e. characterized by a linear scaling of the
mean-square displacement with time), otherwise anomalies (subdiffusive motion)
may occur as discussed in part II.
In the remainder, we consider exclusively the regular case.
Enforcing the Green-Kubo relation \cite{kubo2}, eq. (\ref{eq3_3})
expresses a global condition on the properties
of the particle velocity autocorrelation  tensor
${\bf C}_{vv}(t)= \langle {\bf v}(t) \otimes {\bf v}(0) \rangle_{\rm eq}$,
(where the expected value $\langle \cdot \rangle_{\rm eq}$,  is
referred to the probability mesure of velocity fluctuations at equilibrium),
\begin{equation}
\int_0^\infty {\bf C}_{vv}(t) \, dt = k_B \, T \, \boldsymbol{\eta}_\infty^{-1}
\label{eq3_5}
\end{equation}

This global condition is assumed as the fundamental constraint
in the extension of fluctuation-dissipation theory to correlated
perturbations.

Let $\widehat{\bf C}_{vv}(s)= \int_0^\infty e^{-s \, t} \, {\bf C}_{vv}(t) \, dt$ be the Laplace transform of the velocity autocorrelation tensor.
Equation  (\ref{eq3_3})  corresponds to a constraint on the
value of $\widehat{\bf C}_{vv}(s)$ at $s=0$,
\begin{equation}
\widehat{\bf C}_{vv}(0) = k_B \, T \, \boldsymbol{\eta}_\infty^{-1}
\label{eq3_6}
\end{equation}
All the fluctuational patterns introduced in part I fulfill this
condition.

In dealing with the Fourier extension of the Laplace
transform of autocorrelation functions, say ${\bf C}_{vv}(t)$,
an even extension to negative time is adopted, i.e.
${\bf C}_{vv}(-t)= {\bf C}_{vv}(t)$, $t>0$.

In order to generalize the theory, it is convenient to
consider the functional structures of all the
fluctuation patterns discussed in part I. These patterns
involve white-noise perturbations (either in the form
of distributional derivatives of Wiener processes as in part I,
or distributional derivatives of compound counting processes as in
part II) that are mutually independent of each other.
Let $\boldsymbol{\xi}(t)=(\xi_1(t),\dots, \xi_N(t))$ 
be the vector collecting all these
processes, i.e. in the most general case of ${\mathbb D}{\mathbb I}_{m,E}^{(3,N_d,N_i)}(\boldsymbol{\xi}^{(d)}(t),\boldsymbol{\xi}^{(i)}(t))$,
$\boldsymbol{\xi}(t)=(\boldsymbol{\xi}^{(d)}(t),\boldsymbol{\xi}^{(i)}(t))$,
where  $\boldsymbol{\xi}^{(d)}(t)$ and $\boldsymbol{\xi}^{(i)}(t)$ refer
to the elementary white-noise  forcings acting on the
internal dissipative and fluid-inertial variables (and, due to inertial effects, also
directly to ${\bf v}(t)$, at least as regards $\boldsymbol{\xi}^{(i)}(t)$)
The dimension $N$ of the vector  $\boldsymbol{\xi}(t)$ is $3 \, (N_d+N_i)$,
as we consider the general three-dimensional case.
Independence and white noise nature imply
\begin{equation}
\langle \xi_i(t) \, \xi_j(t^\prime) \rangle = \delta_{ij} \, \delta(t-t^\prime)
\, , \qquad i,j=1,\dots,N
\label{eq3_7}
\end{equation}

Owing to linearity and stationarity of the dynamics associated with 
the fluctuational patterns, the velocity entry $v_h(t)$ can be
expressed as the sum of convolutions of these impulsive stochastic
contributions
\begin{equation}
v_h(t)= \sum_{i=1}^N M_{hi}(t) * \xi_i(t) \, , \qquad h=1,2,3
\label{eq3_8}
\end{equation}
where $M_{hi}(t)$ are the response functions
to the elementary stochastic forcings.
Transforming this equation in the Laplace domain, and setting
$\widehat{v}_h[\omega]= \widehat{v}_h(s)|_{s= \mi \omega}$ 
for the Fourier-Laplace transform of $v_h(t)$ (and similarly
for the other quantties), we have
\begin{equation}
\widehat{v}_h[\omega]= \sum_{i=1}^N \widehat{M}_{hi}[\omega]
\, \widehat{\xi}_i[\omega]
\label{eq3_9}
\end{equation}
Owing to the mutual independence of the elementary stochastic
forcings  eq. (\ref{eq3_7}), we get for the
velocity power spectral density tensor $P_{hk}^{(v)}(\omega)= \langle \widehat{v}_h[\omega] \, \widehat{v}_k^*[\omega] \rangle$, ($z^*$ is the complex conjugate of
the complex number $z$),
\begin{equation}
P_{hk}^{(v)}(\omega)= \sum_{i=1}^N \sum_{j=1}^N
\widehat{M}_{h,i}[\omega] \, \widehat{M}_{k,j}^*[\omega] \, 
P_{ij}^{(\xi)}(\omega)
\label{eq3_10}
\end{equation}
where the power spectral density of the elementary stochastic
forcings equal $P_{ij}^{(\xi)}(\omega)= \delta_{ij}$ by definition
of independent white-noise processes. Thus, eq. (\ref{eq3_10}) reduces
to
\begin{equation}
P_{hk}^{(v)}(\omega)= \sum_{i=1}^N \widehat{M}_{h,i}[\omega] \,
 \widehat{M}_{k,i}^*[\omega] 
\label{eq3_11}
\end{equation}
or in matrix form
\begin{equation}
{\bf P}^{(v)}(\omega)=   \widehat{\bf M}[\omega] \, \widehat{\bf M}^{* T}[\omega] 
\label{eq3_12}
\end{equation}
where  the superscript ``$T$'' indicates transposition.

As developed in part I, the fluctuational patterns satisfy,
by definition, the Kubo fluctuation-dissipation relations of the first and second kind, extended also to the presence of fluid inertial
effects. Therefore, provided that the long-term friction factor
is bounded, they satisfy a-fortiori the global fluctuation-dissipation
relation eq. (\ref{eq3_6}). Enforcing the Wiener-Khinchin
theorem, the velocity  power-spectral density is the Fourier transform
of the velocity autocorrelation function (extended
in an even way over the entire real line, as discussed above). For this reason we
have
\begin{equation}
\widehat{\bf C}_{vv}(0) = \frac{{\bf P}^{(v)}(0)}{2}
\label{eq3_13}
\end{equation}
and the global fluctuation-dissipation relation for
the fluctuational patterns becomes
\begin{equation}
{\bf D}=\widehat{\bf C}_{vv}(0) = \frac{1}{2} \,  \widehat{\bf M}[\omega] \, 
\widehat{\bf M}^{* T}[\omega]
\label{eq3_14}
\end{equation}

Next, consider a generalization of the fluctuational
patterns in which each white-noise elementary stochastic
forcing $\xi_i(t)$ is replaced by a correlated  stationary stochastic
process $q_i(t)$, and these processes are independent of
each other, i.e.
\begin{equation}
\langle q_i(t) \, q_j(t^\prime) \rangle = \delta_{ij} \,
C_{q_i q_j}(|t-t^\prime|) \, , \qquad i,j=1,\dots,N
\label{eq3_15}
\end{equation}
$t, \, t^\prime \in {\mathbb R}$, where $C_{q_i q_i}(|t-t^\prime|)$, $i=1,\dots,N$
 are Lipshitz-continuous functions  of their argument. More precisely
$C_{q_i q_i}(t)$ are smooth for any $|t| >0$, with the only cusp singularity at $t=0$,
where both left and right derivatives exist.
The fluctuation patterns extended in this way to correlated elementary
forcings will be indicated with a letter ``${\mathbb C}$'' preceding the
symbolic code of  the pattern, so that in the simplest
scalar Stokes case ${\mathbb C} {\mathbb S}_{m,E}^{(1,1,0)}(q(t))$ is
the same pattern as ${\mathbb S}_{m,E}^{(1,1,0)}(\xi(t))$ with a correlated
process $q(t)$ substituting $\xi(t)$.

Since the functional structure of a fluctuational pattern remains
unchanged, we still have in the correlated case
\begin{equation}
v_h(t)= \sum_{i=1}^N M_{hi}(t) * q_i(t)
\label{eq3_16}
\end{equation}
Enforcing the mutual independence of the  $q_i(t)$'s, and repeating
the same calculation developed above for the white-noise
case, we arrive to
\begin{equation}
P_{hk}^{(v)}(\omega) = \sum_{i=1}^N \widehat{M}_{hi}[\omega] \, 
\widehat{M}_{k,i}^*[\omega] \, P_{ii}^{(q)}[\omega]
\label{eq3_17}
\end{equation}
Since $P_{ii}^{(q)}[\omega]= \widehat{C}_{q_i q_i}(0)/2$, the
effective diffusivity tensor associated with the correlated pattern is
given by
\begin{equation}
D_{hk}= \sum_{i=1} \widehat{M}_{hi}[0] \, \widehat{M}_{ki}^*[0] \,  \widehat{C}_{q_i q_i}(0)
\label{eq3_18}
\end{equation}
where $\widehat{C}_{q_i q_i}(0)$ is the value of the Laplace transform
of the autocorrelation function $C_{q_i q_i}(t)= \langle q_i(t) \, q_i(0) 
\rangle_{\rm eq}$ of $q_i(t)$. We have the following result: \\

\noindent
{\bf Theorem I} - If the correlated stationary processes $q_i(t)$, $i=1,\dots, N$,
 are
mutually independent, and are characterized by the properties
\begin{equation}
\int_0^\infty C_{q_i q_i}(t) \, d t = \widehat{C}_{q_i q_i}(0) =
\frac{1}{2} \, , \qquad i=1,\dots,N
\label{eq3_19}
\end{equation}
then the correlated fluctuational pattern built upon $\{q_i(t) \}_{i=1}^N$
satisfies the global fluctuation dissipation relation eq. (\ref{eq3_3}). \\

Proof - Under the condition of mutual independence, eq. (\ref{eq3_18})
holds. If the autocorrelation functions of the $q_i(t)$-processes
satisfy eq. (\ref{eq3_19}) then eqs. (\ref{eq3_18}) coincide
with  eq.  (\ref{eq3_14}) for the original fluctuational
pattern in the presence of elementary forcings that corresponds to
the global fluctuation-dissipation relation eq. (\ref{eq3_3}).
This proves the statement. $\diamond$ \\

The condition expressed by eq. (\ref{eq3_19}) admits a simple
physical explanation. Consider a correlated process $q(t)$ satisfying
eq. (\ref{eq3_19}) and characterized by a finite
correlation rate $\nu$, and the diffusive dynamics generated by it,
\begin{equation}
\frac{d x_q(t)}{d t}= q(t)
\label{eqzx1}
\end{equation}
compared with the diffusive dynamics generated by the distributional
derivative $\xi(t)$ of a  Wiener process
\begin{equation}
\frac{d x(t)}{d t} = \xi(t)
\label{eqzx2}
\end{equation} 
starting from $x_q(0)=x(0)=0$. From eq. (\ref{eq3_19}) it follows
that in the long-term $\langle x_q^2(t) \rangle \simeq 2 \, D_q \, t$, where
the diffusivity is $D_q= \widehat{C}_{qq}(0)=1/2$,  and coincides
with that of eq. (\ref{eqzx2}) driven by a Wiener process.
Moreover, in the long term the central limit theorem applies,
and $x_q(t)$ approaches the statistical properties of $x(t)$
defined by eq. (\ref{eqzx2}). This justifies why the
substitution of distributional
derivatives of Wiener processes with correlated processes $q_i(t)$
within the structure of a fluctuational pattern does not
modify the global fluctuation-dissipation  relation,
as the two dynamics possess in the long term  the same statistical
properties.

To complete the analysis, the behavior of the velocity autocorrelation
function at $t=0$ should be investigated, as this
corresponds to the equilibrium values of the second-order
velocity moments. To simplify the analysis consider the scalar case
for a velocity $v(t)$
as the extension to the vectorial problem is straightforward.
Let $C_{vv}(t)$ be the velocity autocorrelation function for any 
fluctuational pattern driven by white-noise elementary forcings,
and $C_{vv, {\rm corr}}(t)$ the  same function for the 
corresponding correlated fluctuational pattern, i.e.
for the same hydrodynamic conditions but with the
elementary forcings expresses as correlated  processes
satisfying the conditions of Theorem I. We have
\begin{equation}
C_{vv}(t)= \langle v^2 \rangle_{\rm eq} \, \phi(t) \, ,
\qquad \langle v^2 \rangle_{\rm eq}= \frac{k_B \, T}{m}
\label{eq3_20}
\end{equation}
and
\begin{equation}
C_{vv, {\rm corr}}(t)= \langle v^2 \rangle_{{\rm eq}, {\rm corr}}
 \, \phi_{\rm corr}(t) 
\label{eq3_21}
\end{equation}
where $\phi(t)$ and $\phi_{\rm corr}(t)$ are the normalized
autocorrelation functions, $\phi(0)=\phi_{\rm corr}(0)=1$.
Since, by Theorem  I the integrals of $C_{vv}(t)$ and $C_{vv, {\rm corr}}(t)$
over $t \in [0,\infty)$ coincide, we have
\begin{equation}
\langle v^2 \rangle_{{\rm eq}, {\rm corr}} = \frac{k_B \, T}{m}
\, \frac{\int_0^\infty \phi(t) \, d t}{\int_0^\infty \phi_{\rm corr}(t)
\, d t}
\label{eq3_22}
\end{equation}

For one and the same structure of the fluctuational pattern, the normalized
velocity autocorrelation  function should be greater in the
presence of correlated forcings than in the case
of $\delta$-correlated   ones, i.e. $\phi_{\rm corr}(t) >\phi(t)$,
and thus from eq. (\ref{eq3_22}) we have
\begin{equation}
\langle v^2 \rangle_{{\rm eq}, {\rm corr}}  \leq \frac{k_B \, T}{m}
\label{eq3_23}
\end{equation}

A proof of the latter statement is provided below.\\

\noindent
{\bf Theorem II} -  For any correlated fluctuational pattern
built upon the correlated processes $\{ q_i(t) \}_{i=1}^N$,
independent of each other and satisfying the conditions of
Theorem I, eq. (\ref{eq3_23}) applies. \\

Proof - Consider a generic velocity entry $v_h(t)$. In the
case of a Wiener-driven fluctuational patter we have from eqs. (\ref{eq3_7})-(\ref{eq3_8})
\begin{eqnarray}
\langle v_h(t) \rangle & = & \sum_{i=1}^N \sum_{j=1}^N \int_0^t M_{hi}(\tau)
\, d \tau \int_0^t M_{hj}(\tau^\prime) \langle \xi_i(t-\tau) \, \xi_j(t-\tau^\prime) \rangle \, d \tau^\prime \nonumber \\
& = & \sum_{i=1}^N \int_0^t M_{hi}^2(\tau) \, d \tau
\label{eqzc1}
\end{eqnarray}
the equilibrium value corresponds to the limit for
$t \rightarrow \infty$, and fulfills eq. (\ref{eq2}),
\begin{equation}
\langle v_h^2 \rangle_{\rm eq}= \sum_{i=1}^N \int_0^\infty M_{hi}^2(\tau)
\, d \tau = \frac{k_B \, T}{m}
\label{eqzc2}
\end{equation}
Next, consider its correlated counterpart substituting $\{ \xi_i(t) \}_{i=1}^N$ with
independent correlated processes $\{ q_i(t) \}_{i=1}^N$.
Enforcing their mutual independence we have
\begin{eqnarray}
\langle v_h^2(t) \rangle_{\rm corr} & = &
\sum_{i=1}^N \int_0^t M_{hi}(\tau) \, d \tau \int_0^t M_{hi}(\tau^\prime)
\, C_{q_i q_i}(|\tau- \tau^\prime|) \, d \tau^\prime \nonumber \\
& = &  2 \, \sum_{i=1}^N  \int_0^t M_{hi}(\tau^\prime) \, d \tau^\prime
\int_{\tau^\prime}^t M_{hi}(\tau) \, C_{q_i q_i}(\tau- \tau^\prime) \, d \tau
\label{eqzc3}
\end{eqnarray}
In the limit for $t \rightarrow \infty$ the equilibrium value is achieved,
\begin{eqnarray}
\langle v_h^2(t) \rangle_{{\rm eq}, {\rm corr}} & = &  2 \, \sum_{i=1}^N
\int_0^\infty M_{hi}(\tau^\prime) \, d \tau^\prime
\int_{\tau^\prime}^\infty M_{hi}(\tau) \,
C_{q_i q_i}(\tau- \tau^\prime) \, d \tau \nonumber \\
& = & 2 \, \sum_{i=1}^N
\int_0^\infty M_{hi}(\tau^\prime) \, d \tau^\prime
\int_0^\infty M_{hi}(\theta + \tau^\prime) \, C_{q_i q_i}(\theta) \, d \theta
\nonumber \\
& = & 2 \, \sum_{i=1}^N \int_0^\infty C_{q_i q_i}(\theta) \, d \theta
\int_0^\infty M_{hi}(\tau^\prime) \, M_{hi}(\theta+\tau^\prime) \, 
d \tau^\prime
\label{eqzc4}
\end{eqnarray}
In the last integral, the Schwartz inequality can be applied,
namely
\begin{equation}
\left | \int_0^\infty M_{hi}(\tau^\prime) \, M_{hi}(\theta+\tau^\prime) \, 
d \tau^\prime \right | \leq \int_0^\infty M_{h,i}^2(\tau^\prime) \, d \tau^\prime 
\label{eqzc5}
\end{equation}
Thus,
\begin{eqnarray}
\langle v^2_h \rangle_{{\rm eq}, {\rm corr}} & \leq  & 
\sum_{i=1}^N   \left | \, 2 \, \int_0^\infty
C_{q_i  q_i}(\theta) \, d \theta \, \right | \, \left | \int_0^\infty M_{hi}^2(\tau^\prime) \, d \tau^\prime \, \right | \nonumber \\
& \leq & \sum_{i=1}^N \int_0^\infty M_{hi}^2(\tau^\prime) \, d \tau^\prime  = \frac{k_B \, T}{m}
\label{eqzc6}
\end{eqnarray}
where we have used eq. (\ref{eq3_19}) and  eq. (\ref{eqzc2}).$\diamond$.\\

The proof of Theorem II is grounded on the Schwarz inequality
eq. (\ref{eqzc5}),in which the equal sign applies only
in the limit case $q_i(t)$  tending towards $\xi_i(t)$. This
is interesting as it indicates that a non-vanishing added
mass, no matter how small, characterized any correlated
fluctuational patterns. If the elementary stochastic
forcings in a fluctuational pattern possess some
characteristic correlation time, either
the global fluctuation-dissipation relation (as in the
present formulation of the theory) or the
equipartition result can be fulfilled, but never both
at the same time. The only case in which these
two properties are simultaneously satisfied occurs
for the classical fluctuational patterns discussed
in part I and II driven by white-noise processes.

Eq. (\ref{eq3_23}) indicates that an added mass 
arises as an emergent property of correlated fluctuations  defining
the elementary stochastic forcing of a correlated fluctuational pattern (the ${\mathbb C}$-patterns).

In a physical perspective, this results applies on equal footing
to incompressible and compressible flows, as it is not
a property  of the hydromechanic kernels, but a consequence
of the regular nature of the hydrodynamic fluctuations.
A connection between this fluctuational property and the 
fluid-particle hydromechanics is addressed in Section \ref{sec6}
via the   fluctuation-inertia relation. Some
prototypical examples of application of Theorem II are discussed
in the next Section.

\section{Examples}
\label{sec4}

This Section discusses several prototypical examples, considering
for notational simplicity the scalar case, i.e. a scalar velocity $v(t)$,
corresponding to the isotropic dynamics of a spherical object.

Before analyzing some case studies, let us consider how to
construct admissible correlated processes, i.e. stochastic 
processes $q(t)$ fulfilling eq. (\ref{eq3_19}).
There are two simple ways of defining them: either 
through successive low-pass filterings of distributional
derivatives of Wiener processes or by considering discontinuous
correlated processes, such as the Poisson-Kac processes \cite{kac,pk1,pk2}.

Consider the first case. The simplest example is the output $q_\nu(t)$
of a low-pass
filtering of $\xi(t)$, i.e. of a distributional derivative of a Wiener
process  through a filter with characteristic time $\nu^{-1}$
under the condition that, for  $\nu \rightarrow \infty$ $q_\nu(t)$, converges
stochastically to $\xi(t)$. 
 For notational simplicity we set $q(t)=q_\nu(t)$.
The process is defined by the dynamic equation
\begin{equation}
\frac{d q(t)}{ d t}= - \nu \, (q(t)- \xi(t))
\label{eq4_1}
\end{equation}
which is characterized by the equilibrium (long-term) properties: $\langle q^2 \rangle_{\rm eq}=\nu/2$ and 
\begin{equation}
C_{qq}(t)= \frac{ \nu \, e^{- \nu \, t}}{2}
\label{eq4_2}
\end{equation}
where $C_{qq}(t) = \langle q(t) \, q(0) \rangle_{\rm eq}$
is its autocorrelation function. Therefore, $\widehat{C}_{qq}(0)=1/2$, consistently with the
condition expressed by Theorem I. 

Similarly, this mechanism can be generalized to
a cascade of low-pass filters  of different characteristic times.
For example, in the case of two filters we have
for $q(t)$
\begin{eqnarray}
\frac {d q(t)}{d t} & = & - \nu (q(t)- q_1(t)) \nonumber \\
\frac{d q_1(t)}{d t} & = &- \nu_1 (q_1(t) - \xi(t))
\label{eq4_3}
\end{eqnarray}
and it is characterized by the steady-state moments
\begin{equation}
m_{q_1 q_1}^* = \langle q_1^2 \rangle_{\rm eq}= \frac{\nu_1}{2} \, ,
\qquad m_{q q_1}^*=  \langle  q \, q_1 \rangle_{\rm eq} = m_{qq}^*=\langle q^2 
\rangle_{\rm eq} = \frac{ \nu \, \nu_1}{2 (\nu+ \nu_1)}
\label{eq4_4}
\end{equation}
Its autocorrelation function $C_{qq}(t)$  is the solution of the
equations
\begin{eqnarray}
\frac{d C_{qq}(t)}{d t} & = & - \nu ( C_{qq}(t) - C_{q_1 q}(t)) \nonumber \\
\frac{d C_{q_1 q}(t)}{d t} & = & - \nu_1 C_{q_1 q}(t)
\label{eq4_5}
\end{eqnarray}
equipped with the initial conditions $C_{qq}(0)=m_{qq}^*$, $C_{q_1 q}(0)= m_{q q_1}^*$, the solution of which  reads
in the Laplace domain
\begin{equation}
\widehat{C}_{qq}(s)= \frac{m_{qq}^*}{s+ \nu} + \frac{\nu \, m_{q q_1}^*}{(s+\nu) (s+\nu_1)}
\label{eq4_6}
\end{equation}
Substituting the values of the stationary moments
eq. (\ref{eq4_4}) into eq. (\ref{eq4_6}) ,   $\widehat{C}_{qq}(0)=1/2$ follows, confirming
its admissibility.

Alternatively, one may define $q(t)$ as a Poisson-Kac process
or its generalizations \cite{kac,pk1,pk2}. The simplest example is
provided by the process
\begin{equation}
q(t)= c \, (-1)^{\chi(t,\nu)}
\label{eq4_7}
\end{equation}
where $\chi(t,\nu)$ is a Poisson process with rate $\nu$, and $c$ a constant
to be determined.
Since, for $t \geq 0$, we have
\begin{equation}
C_{qq}(t)= c^2 \, \left \langle (-1)^{\chi(t,\nu)} \, (-1)^{\chi(0,\nu)}
\right \rangle = c^2 \, e^{-2 \, \nu \, t}  \; \; \Rightarrow \; \;
\int_0^\infty C_{qq}(t) \, d t= \frac{c^2}{2 \, \nu}
\label{eq4_8}
\end{equation}
the admissibility of this process, i.e. the condition of Theorem I, implies $c = \sqrt{\nu}$.

Next, consider the application of correlated forcings to the
prototypical fluctuational patterns. The process $q(t)$ defined
by eq. (\ref{eq4_1}) will be explicitly considered.

\subsection{${\mathbb C} {\mathbb S}_{m,E}^{(1,1,0)}(q(t))$}
\label{subsec41}

\begin{figure}
\includegraphics[width=10cm]{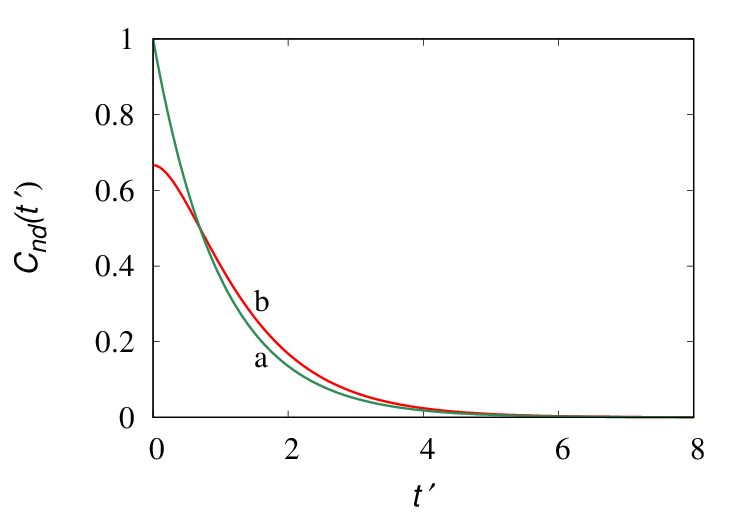}
\caption{Normalized velocity autocorrelation functions $C_{\rm nd}(t^\prime)$ vs the nondimensional
time $t^\prime$. Line (a) refers to ${\mathbb S}_{m,E}^{(1,1,0)}(\xi(t))$, line (b)
to ${\mathbb C} {\mathbb S}_{m,E}^{(1,1,0)}(q(t))$ at a value of $\nu^*=2$.}
\label{Fig1}
\end{figure}

For the  scalar Stokesian pattern we have $a=\sqrt{k_B \, T \, \eta}$, and the dynamics of ${\mathbb C} {\mathbb S}_{m,R}^{(1,1,0)}(q(t))$
is thus defined by
\begin{eqnarray}
m \, \frac{d v(t)}{d t} & = & - \eta \, v(t) +  \sqrt{2} \, a \, q(t) \nonumber \\
\frac{d q(t)}{d t} & = & - \nu (q(t)-\xi(t))
\label{eq4_9}
\end{eqnarray}
Setting $\alpha= \eta/m$, $\beta= \sqrt{2} a /m$, the
second-order moments $m_{vv}(t)=\langle v^2(t) \rangle$,  $m_{vv}(t)=\langle v(t)  q(t) \rangle$, $m_{qq}(t)=\langle q^2(t) \rangle$ evolve according
to the dynamics
\begin{eqnarray}
\frac{d m_{vv}(t)}{d t} & = & - 2 \, \alpha \, m_{vv}(t) + 2 \beta \, m_{v q}(t)
\nonumber \\
\frac{d m_{vq}(t)}{d t} & = & - \alpha \, m_{vq}(t) + \beta \, m_{qq}(t) -
\nu \, m_{vq}(t) 
\label{eq4_10} \\
\frac{d m_{qq}(t)}{d t} &  = & - 2 \,  \nu m_{qq}(t)+ \nu^2
\nonumber
\end{eqnarray}
providing the steady-state values
\begin{equation}
m_{vv}^*= \frac{\beta^2 \, \nu}{2 \, \alpha \, (\alpha+\nu)} \, , \qquad
m_{vq}^*= \frac{ \beta \, \nu}{2 \,  (\alpha+\nu)} \, ,
\quad m_{qq}^*= \frac{\nu}{2}
\label{eq4_11}
\end{equation}
The  squared velocity variance at equilibrium is given by
\begin{equation} 
\langle v^2 \rangle_{\rm eq}= \frac{k_B \, T}{m \, (1+ \tau_c^*)} 
\label{eq4_12}
\end{equation}
where $\tau_c^*$  is the dimensionless correlation
time equal to the ratio of the  correlation  time $1/\nu$ to the
dissipation   time $m/\eta$,
\begin{equation}
\tau_c^*= \frac{1}{\nu^*}=  \frac{\eta}{m \, \nu}
\label{eq4_13}
\end{equation}
where $\nu^*=\nu/\alpha$.
It follows from eq. (\ref{eq4_12}) that the nondimensional
added mass of this pattern is given by
\begin{equation}
\frac{m_a}{m}= \tau_c^*
\label{eq4_14}
\end{equation}
As  regards the velocity autocorrelation function $C_{vv}(t)$, we have
\begin{eqnarray}
\frac{d C_{vv}(t)}{d t} & = & - \alpha \, C_{vv}(t) + \beta \, C_{qv}(t)
\nonumber \\
\frac{d C_{qv}(t)}{ d t}  & = & - \mu \, C_{q v}(t)
\label{eq4_15}
\end{eqnarray}
equipped with the initial conditions
$C_{vv}(0)=m_{vv}^*$, $C_{qv}(0)=m_{vq}^*$, The Laplace trasform  of $C_{vv}(t)$,
solution of eq. (\ref{eq4_15}), is
\begin{equation}
\widehat{C}_{vv}(s)= \frac{m_{vv}^*}{s+\alpha} + \frac{\beta \, m_{vq}^*}{(s+\alpha) (s+\nu)}
\label{eq4_16}
\end{equation}
from which, substituting the equilibrium values for the moments eq. (\ref{eq4_11}), the Stokes-Einstein relation $\widehat{C}_{vv}(0)= k_B \, T/m \,
\eta$ is recovered. The explicit expression for  the nondimensional
velocity autocorrelation $C_{\rm nd}(t^\prime) = m \, C_{vv}(t)/k_B \, T
|_{t=t^\prime/\alpha}$ reads for $t>0$
\begin{equation}
C_{\rm nd}(t^\prime)=
\left \{
\begin{array}{lll}
\frac{1}{1+1/\nu^*}  e^{-t^\prime}+ \frac{\nu^*}{1-{\nu^*}^2} \left [ e^{-\nu^* \,
t^\prime} - e^{-t^\prime} \right ] & \;\;\; & \nu^* \neq 1 \\
\frac{1}{1+1/\nu^*}  e^{-t^\prime} + \frac{\nu^*}{1+\nu^*} \, t^\prime \, e^{-t^\prime} & & \nu^*=1
\end{array}
\right . 
\label{eq4_17}
\end{equation}
where $t^\prime=\alpha \, t$ is the nondimensional time rescaled with respect to
the dissipation time.
As a realistic  situation, consider  the case where the fluctational
added mass coincides with  the hydrodynamic added mass for an
object possessing the same density of the solvent fluid. In this case
$m_a/m=1/2$, and this corresponds to $\nu^*=2$. Figure \ref{Fig1}
depicts the  comparison of the  normalized velocity autocorrelation functions  for
the patterns ${\mathbb S}_{m,E}^{(1,1,0)}(\xi(t))$ and ${\mathbb C} {\mathbb S}_{m,E}^{(1,1,0)}(q(t))$
at $\nu^*=2$.  A slight difference can be observed 
between the $\delta$-correlated and the correlated patterns.

As a final observation, consider the Langevin condition. In this model,
$R(t)$ is proportional to $q(t)$, and we have ($t \geq 0$)
\begin{equation}
\langle R(t) \, v(0) \rangle_{\rm eq} = \sqrt{2 \, k_B \, T \, \eta} \,
\langle q(t) \, v(0) \rangle_{\rm eq} =
\frac{k_B \, T \, \nu}{1+ \nu \, m/\eta} \, e^{-\nu \, t}
\label{eq4_18}
\end{equation}
Thus, for any finite $\nu$, the Langevin condition is not
satisfied, as intuitively expected. In the limit for $\nu \rightarrow \infty$,
i.e. when $q(t)$ approaches $\xi(t)$,
one gets from eq. (\ref{eq4_18}) 
\begin{equation}
\lim_{\nu \rightarrow \infty}  \langle R(t) \, v(0) \rangle_{\rm eq} =
\left \{
\begin{array}{lll}
k_B \, T \, \eta/m & \;\; \; & t=0 \\
0 & & t>0 
\end{array}
\right .
\label{eq4_19}
\end{equation}
The  correlation  between $R(t)$ and $v(0)$ is vanishing
for any $t>0$, and attains at $t=0$  a finite value. So for any 
practical purpose the limit eq. (\ref{eq4_19}) converges in  the L$^2$-norm to
the Langevin condition, i.e. to zero.

\subsection{${\mathbb C} {\mathbb D}_{m,E}^{(1,1,0)}(q(t))$}
\label{subsec4_2}

\begin{figure}
\includegraphics[width=10cm]{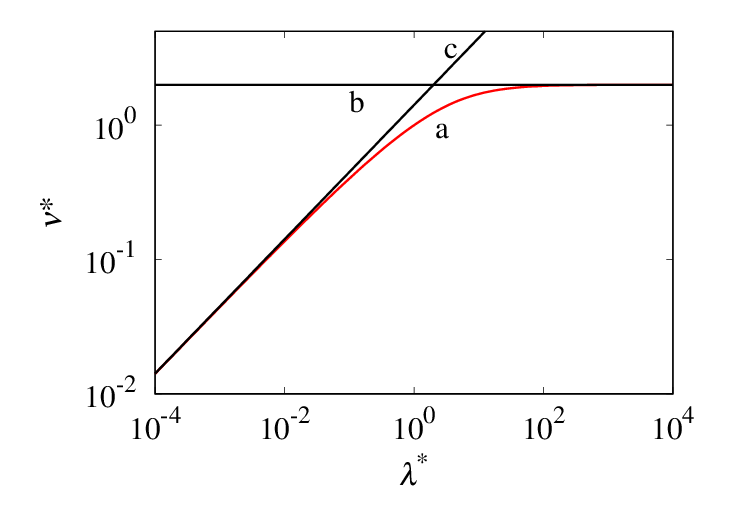}
\caption{$\nu^*$ vs $\lambda^*$ (line a) for the ${\mathbb C}  {\mathbb D}_{m,E}^{(1,1,0)}(q(t))$ pattern
providing $m_a/m=1/2$. Line (b) refers to the Stokesian limit $\nu^*=2$, line (c) to the
low-relaxation rate limit $\nu^* = \sqrt{2 \, \lambda^*} $.}
\label{Fig2}
\end{figure}

The next pattern in order of complexity is
the correlated dissipative pattern   ${\mathbb C} {\mathbb D}_{m,E}^{(1,1,0)}(q(t))$   associated
with the pattern ${\mathbb D}_{m,E}^{(1,1,0)}(\xi(t))$,
in the presence of a single exponential memory term.  
This correlated pattern is defined by
the dynamics
\begin{eqnarray}
\frac{d v(t)}{ d t}& - & \alpha \, \theta(t) \nonumber \\
\frac{d \theta(t)}{d t} & = &- \lambda \, \theta(t) + \lambda \, v(t)
+ \sqrt{2} \, b \, \lambda \, q(t)
\label{eq4_20} \\
\frac{d q(t)}{d t} & = & -  \nu (q(t)- \xi(t))
\nonumber
\end{eqnarray}
where $b=\sqrt{k_B \, T/\eta}$.
The steady-state moments are given by $m_{v \theta}^*=0$, $m_{qq}^*=\nu/2$,
\begin{equation}
m_{\theta q}^*=  \frac{b \, \lambda \, \nu}{ \sqrt{2} \, (\lambda + \nu
+ \alpha \, \lambda/\nu)} \, , \qquad
m_{vq}^* = -\frac{\alpha \, b \, \lambda}{\sqrt{2} \, (\lambda + \nu
+ \alpha \, \lambda/\nu)} \, , \quad
m_{\theta \theta}^* = \frac{b^2 \, \lambda \, \nu}{(\lambda + \nu
+ \alpha \, \lambda/\nu)}
\label{eq4_21}
\end{equation}
and
\begin{equation}
m_{vv}^*= \langle v^2 \rangle_{\rm eq} =
\frac{ b^2 \, \alpha \, (\lambda+\nu)}{(\lambda + \nu
+ \alpha \, \lambda/\nu)}
= \frac{k_B \, T}{m} \left (1+ \frac{\alpha \, \lambda}{\nu \, (\lambda + \nu)}
\right )^{-1}
\label{eq4_22}
\end{equation}
and this corresponds to a fluctuational added mass
\begin{equation}
\frac{m_a}{m}= \frac{\eta \, \lambda}{m \, \nu \, (\lambda + \nu)}
\label{eq4_23}
\end{equation}
Also for this pattern, consider the  case $m_a/m=1/2$. This corresponds
to a dimensionless correlation rate $\nu^*=\nu/\alpha$, depending
on the dimensionless relaxation rate $\lambda^*=\lambda/\alpha$,  
given by
\begin{equation}
\nu^*= \frac{\lambda^*}{2} \left [
\sqrt{1 + \frac{8}{\lambda^*}} -1 \right ]
\label{eq4_24}
\end{equation}
Figure \ref{Fig2} depicts the behaviour of $\nu^*$, providing an added
mass equal half of the particle mass, as a function of $\lambda^*$.

For very large $\lambda^*$, one recovers the Stokesian result discussed
above $\nu^* \simeq 2$. For $\lambda^* \ll 1$, $\nu^*  \sim \sqrt{2 \, \lambda^*}$.
This result is not surprising. In a viscoelastic fluids, the
correlation rate $\nu$ to achieve a given added mass depends on
the viscoelastic relaxation rate $\lambda$, and if $\lambda \ll \alpha$,
$\nu  \simeq  \sqrt{ 2 \, \lambda \, \alpha}$, i.e. it is proportional to the
geometric mean of the dissipation and relaxation rates.

\begin{figure}
\includegraphics[width=10cm]{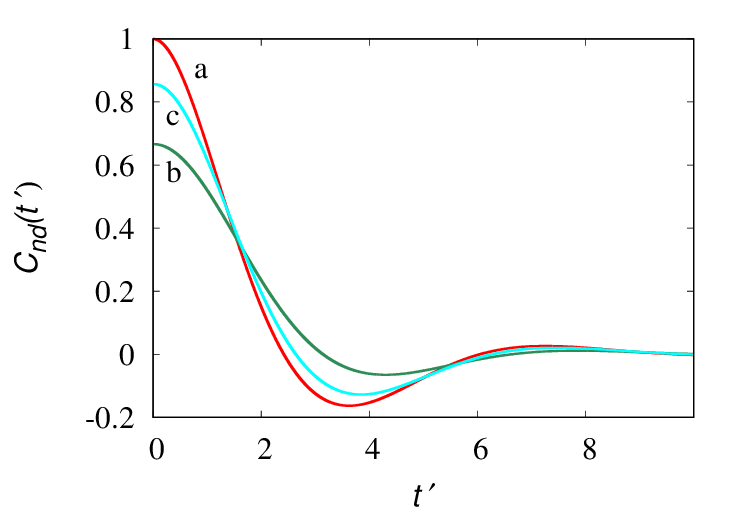}
\caption{$C_{nd}(t^\prime)$ vs $t^\prime$ for the ${\mathbb C} {\mathbb D}$-patterns at $\lambda^*=1$.
Line (a) refers to the corresponding ${\mathbb D}$-pattern. Line (b) refers to $\nu^*=1$ corresponding
to $m_a/m=1/2$, line (c) to $\nu^*=2$ corresponding to $m/m_a=1/6$.} 
\label{Fig3}
\end{figure}

Next, consider the  velocity autocorrrelation function.
Setting $C_{nd}(t^\prime)= m \, C_{vv}(t^\prime)/k_B T$
where $t^\prime=\alpha t$ is the nondimensional time,
and $\widetilde{C}_{\theta v}(t^\prime)= C_{\theta v}(t^\prime)/b^2 \alpha$,
$\widetilde{C}_{qv}(t^\prime)= \sqrt{2} \, C_{qv}(t^\prime)/b \alpha$,
the nondimensional velocity autocorrelation function is
the solution of this system of equations
\begin{eqnarray} 
\frac{d C_{nd}(t^\prime)}{d t^\prime} & = &
- \widetilde{C}_{\theta v}(t^\prime) \nonumber \\
\frac{d \widetilde{C}_{\theta v}(t^\prime)}{d t^\prime}
& = & - \lambda^* \, \widetilde{C}_{\theta v}(t^\prime) +
\lambda^* \, C_{nd}(t^\prime) + \lambda^* \, \widetilde{C}_{q v}(t^\prime)
\label{eq4_25} \\
\frac{d \widetilde{C}_{qv}(t^\prime)}{d t^\prime}
& = & - \nu^* \widetilde{C}_{qv}(t^\prime)
\nonumber
\end{eqnarray}
equipped with the initial conditions
$C_{nd}(0)= (1+\lambda^*/ \nu^*(\lambda^*+\nu^*))^{-1}$, 
$\widetilde{C}_{\theta v}(0)=0$, 
$\widetilde{C}_{q v}(0)= -\lambda^*/(\lambda^*+\nu^*+\lambda^*/\nu^*)$.
Figure \ref{Fig3} depicts the behaviour of the normalized velocity
autocorrelation function at $\lambda^*=1$, for values of $\nu^*$ providing $m_a/m=1/2$
and $m_a/m=1/6$ respectively.
$\lambda^*$.

To conclude, consider the Langevin condition. Since
$R(t)= -  m \, \sqrt{2} \, \alpha  \, b \, \lambda \, e^{-\lambda \, t} * q(t)$
we have
\begin{equation}
\langle R(t) \, v(0) \rangle_{\rm eq} = -  m \, \sqrt{2} \, \alpha  \, b \, \lambda \,  \left (e^{-\lambda \, t} * C_{qv}(t)  \right ) =
-  m \, \sqrt{2} \, \alpha  \, b \, \lambda \, m_{v q} \, 
\left ( e^{-\lambda \, t}  * e^{-\nu \,  t} \right )
\label{eq4_27}
\end{equation}
Expressing the Langevin term with respect to the dimensionless  time $t^\prime$,
eq. (\ref{eq4_27}) becomes
\begin{equation}
\langle R(t^\prime) \, v(0) \rangle_{\rm eq} = \frac{k_B \, T \, \eta}{m}
\, \frac{ (\lambda^*)^2}{\lambda^* + \nu^* +  \lambda^*/\nu^*}
\,
\left \{
\begin{array}{lll}
\frac{e^{-\nu^* t^\prime}-e^{-\lambda^*  t^\prime}}{\lambda^*-\nu^*} &
\;\;\; \nu \neq \lambda  \\
t^\prime \, e^{-\lambda^*  t^\prime} & & \nu= \lambda
\end{array}
\right .
\label{eq4_28}
\end{equation}
In this case, $\langle R(0) \, v(0) \rangle_{\rm eq} =0$,
while $\langle R(t^\prime) \, v(0) \rangle_{\rm eq}>0$ for $t^\prime>0$.

\subsection{${\mathbb C} {\mathbb I}_{m,E}^{(1,1,1)}(q(t), q_1(t))$}
\label{subsec4_3}

The next prototypical model of increasing hydrodynamic complexity involves the occurrence of a fluid inertial
contribution superimposed to a Stokesian dissipative term,
i.e. ${\mathbb I}_{m,E}^{(1,1,1)}(\xi(t), \xi_1(t))$.
In the presence of correlated forcing it becomes
${\mathbb C} {\mathbb I}_{m,E}^{(1,1,1)}(q(t), q_1(t))$
where  $q(t)$ and $q_1(t)$  are the low-pass filtering of the 
white-noise independent processes $\xi(t)$ and $\xi_1(t)$.
The dynamics of this correlated pattern reads
\begin{eqnarray}
\frac{d v(t)}{d t} & = & -(\gamma+\alpha) \, v(t) + \gamma \, z(t)
+ \sqrt{2} a \ \, q(t) + \sqrt{2} \, d \, q_1(t) \nonumber \\
\frac{d z(t)}{d t} & = & -\mu  z(t) + \mu \, v(t) + \sqrt{2} \, c \, q_1(t)
\label{eq4_29} \\
\frac{d q(t)}{d t} & = & - \nu (q(t)-\xi(t)) \nonumber \\
\frac{d q_1(t)}{d t} & = & - \nu_1 (q(t)-\xi_1(t)) \nonumber 
\end{eqnarray}
where $\alpha=\eta/m$ , $\gamma=g/m$ and (see part I)
\begin{equation}
a= \delta = \frac{\sqrt{k_B \, T \, \eta}}{m} \, ,
\quad
d = \sqrt{\frac{\gamma}{\alpha}} \, \delta \, , \quad
c = - \frac{\delta}{\sqrt{\alpha \, \gamma}}
\label{eq4_30}
\end{equation}
From the analysis of the moments at steady state, one
immediately gets $m_{qq}^*=\nu/2$, $m_{q_1 q_1}^*=\nu_1/2$, $m_{q q_1}^*=0$.
The remaining seven moments satisfy a linear system, the
solution of which provides the value for
 $m_{vv}^*=\langle v^2 \rangle_{\rm eq}$, see the Appendix.
For finite values of  $\nu$ and $\nu_1$, we have 
$\langle v^2 \rangle_{\rm eq} < k_B T/m$, 
and
\begin{equation}
\lim_{\nu \rightarrow \infty} \lim_{\nu_1 \rightarrow \infty}
\frac{m \, \langle v^2 \rangle_{\rm eq}}{k_B \, T} =1
\label{eq4_31}
\end{equation}
consistently with the limit of $q(t)$ and $q_1(t)$ tending to
white-noise processes.
From the analysis of the velocity autocorrelation function it follows that
the global
fluctuation-dissipation relation is satisfied for any choice of $\nu$ and $\nu_1$, as expected
from Theorem I. 
\begin{figure}
\includegraphics[width=10cm]{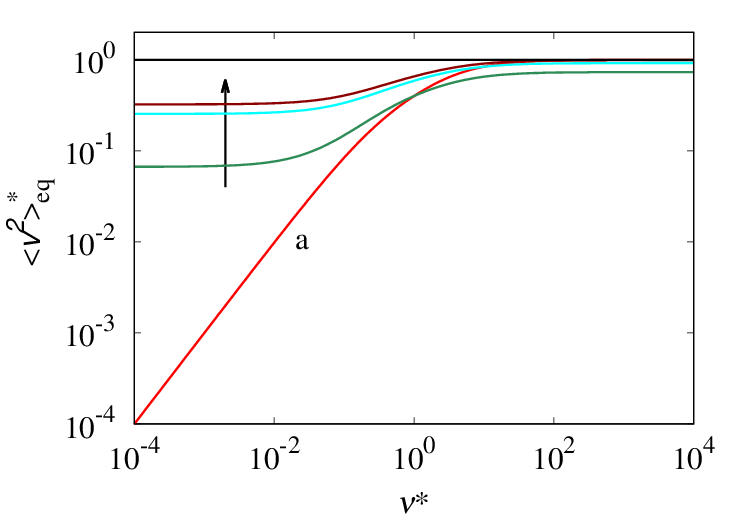}
\caption{Normalized squared velocity variance $\langle v^2 \rangle_{\rm eq}^*$ vs $\nu^*$  at $\mu=1$, $\gamma=1$, for different values of the rate $\nu_1^*$.
Line (a) refers to $\nu_1=\nu^*$. The remaining lines indicated by
an arrow refer to $\nu=1$ for increasing values of $\nu_1^*=1,\,10,\,100$. The horizontal line
depicts the limit behaviour $\langle v^2 \rangle_{\rm eq}^*=1$.}
\label{Fig4}
\end{figure}
Figure \ref{Fig4} depicts the behaviour of the normalized squared velocity variance $\langle v^2 \rangle_{\rm eq}^*=
m \, \langle v^2 \rangle_{\rm eq}/k_B \, T$ obtained from 
${\mathbb C} {\mathbb I}_{m,E}^{(1,1,1)}(q(t), q_1(t))$ patterns as a function of $\nu^*=\nu/\alpha$, for different
settings of $\nu_1^*=\nu_1/\alpha$.

\section{Regularity of  the velocity fluctuations}
\label{sec5}

There is another general qualitative property that arises as a consequence of
fluid-inertial interactions, related to the regularity of velocity fluctuations,
in the meaning of their H\"older continuity \cite{tricot}.
Consider a single realization of the stochastic velocity process $v(t)$ in
some  time interval  $I=[a,b]$, $a>0$, $b>a$.
The  process is H\"older continuous in $I$ with exponent $H$ if, for any generic realization
$v(t)$, and for any
$t_1,t_2 \in I$, $t_2>t_1$,
, there exist constants $C_1,C_2>0$, $C_2>C_1$ such that 
\begin{equation}
C_1 \, |t_2 - t_1 |^H \leq | v(t_2) - v(t_1) | \leq C_2 \,  |t_2 - t_1 |^H 
\label{eqreg1}
\end{equation}
The exponent $H$ is referred to as the H\"older exponent. If $H=1$,  $v(t)$ is Lipshitz continuous
(something more than continuity, something less than differentiability). Values of $H$
less than $1$ implies that the function $v(t)$, (regarding a realization of the process as a function of time $t$),
displays a singular  behaviour almost everywhere in  $I$, in the meaning that it cannot
be defined a tangent vector at almost any $t \in I$. This  follows readily 
for the simple observation that, setting $t_2=t_1+\Delta t$,
$\Delta t>0$, eq. (\ref{eqreg1}) implies
\begin{equation}
\frac{C_1}{\Delta t^{1-H}} \leq \frac{|v(t_1+\Delta t)-v(t_1)|}{\Delta t} \leq \frac{C_2}{\Delta t^{1-H}} 
\label{eqreg2}
\end{equation}
Therefore, in the limit for $\Delta t \rightarrow 0$, the absolute value of the derivative of $v(t)$ at $t_1$
diverges. In the case of a realization of a Wiener process $w(t)$, it follows from its definition \cite{lasota} that $w(t_2)-w(t_1) = \sqrt{t_2-t_1} \,
r$ where  $r \in {\mathcal N}(0,1)$ is a random variable normally distributed, and thus  $H=1/2$ \cite{falconer}.

An H\"older exponent less than $1$ is the manifestation of the fractal nature ofa generic 
realization $v(t)$. More precisely, the fractal dimension $d_v$ of  $v(t)$, 
characterized by the
H\"older exponent $H$,  is given by  \cite{tricot}
\begin{equation}
d_v=2 - H
\label{eqreg3}
\end{equation}
For a Wiener process, its fractal dimension equals $3/2$. Eq. (\ref{eqreg3}) implies that the estimate of $H$ can be 
conveniently obtained by considering the scaling of the length $L(\Delta t)$ of $v(t)$ in the interval $I$ as a function of the
temporal resolution $\Delta t$, since
\begin{equation}
L(\Delta t) \sim \Delta t^{1-H}
\label{eqreg4}
\end{equation}
The length-scale analysis associated with eq. (\ref{eqreg4})
is experimentally feasible and computationally very simple as it requires solely
a single, statistically representative, realization of the particle velocity, i.e. a single time series.

Next, consider the
different hydromechanic mechanisms and their influence as regards the regularity properties of the
particle velocity $v(t)$.
The  structure of the different
fluctuational patterns introduced in part I as regards the action of the stochastic forcings is  particularly useful in this analysis.
Without loss of generality we consider  the onedimensional  scalar case (spherical particle of radius $R_p$)
and the nondimensional formulation adopted  in part I, rescaling time with respect to the characteristic dissipation
time and  velocity so that $\langle v^2 \rangle_{\rm eq}=1$.
In order to consider a model closely connected with the high-resolution experiments \cite{exp1,exp2,exp3,exp4}, we include 
also the influence of a harmonic potential ensuring  in the experiments particle localization due to  the action of an optical
trap.

\begin{figure}
\includegraphics[width=10cm]{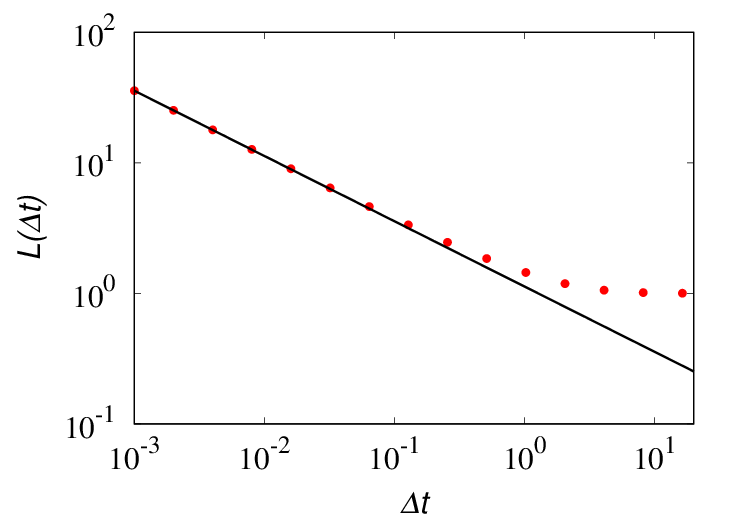}
\caption{Length-resolution analysis, $L(\Delta t)$ vs $\Delta t$, for a realization of
the ${\mathbb S}_{1,1}^{(1,1,0)}(\xi(t))$ pattern in the presence of a harmonic potential with $k=0.2$.
Symbols ($\bullet$) are the results of the stochastic simulations, the solid line represents the
scaling $L(\Delta t) \sim \Delta t^{-1/2}$.}
\label{Fig5}
\end{figure}

To begin with, consider the Stokes approximation, i.e. the pattern ${\mathbb S}_{1,1}^{(1,1,0)}(\xi(t))$, the
dynamics of which, in the presence of a quadratic potential, reads
\begin{eqnarray}
\dot{x}(t) & = & v(t) \\
\dot{v}(t) & = & - v(t) - k \, x(t) +\sqrt{2} \, \xi(t)
\label{eqreg5}
\end{eqnarray}
where $\dot{x}(t)=d x(t)/dt$ and similarly for the other variables, and $k$ is the nondimensional spring constant of the trap.
It is a typical feature of the Stokesian pattern that $\dot{v}(t)$ is proportional to the
distributional derivative $\xi(t)=d w(t)/d t$ of a Wiener process $w(t)$ and thus 
$v(t) \sim w(t)$. Consequently, the regularity properties of $v(t)$ stemming from ${\mathbb S}_{1,1}^{(1,1,0)}(\xi(t))$
are those of a Wiener process, implying the fractal nature of the velocity  realizations with a H\"older exponent 
$H=1/2$.
Figure \ref{Fig5} depicts the scaling of the length $L(\Delta t)$ vs the temporal resolution $\Delta t$  for a velocity
realization of this pattern. 

The subsequent pattern in order of hydrodynamic complexity is ${\mathbb D}_{1,1}^{(1,1,0)}(\xi^{(d)}(t))$ 
characterized by the presence of a dissipative memory dynamics. We consider for simplicity  the case of a single mode, but the
analysis applies to a generic superposition of dissipative modes.
The equations of motion of this pattern are
\begin{eqnarray}
\dot{x}(t) & = & v(t) \nonumber \\
\dot{v}(t)  & = & - \lambda \, \theta(t) - k \, x(t)  
\label{eqreg6} \\
\dot{\theta}(t) & = & -\lambda \, \theta(t)+ v(t) + \sqrt{2} \, \xi^{(d)}(t)
\nonumber
\end{eqnarray}
where $\xi^{(d)}(t) = d w^{(d)}(t)/dt$, and $w^{(d)}(t)$ is a Wiener process.
The  characteristic structure of the  ${\mathbb D}$-patterns is that the Wiener forcings, in the
present case $\xi^{(d)}(t)$ do not act directly on the velocity dynamics, but solely on the
auxiliary $\theta$-variables. Since $\theta(t)$ is
continuous and almost everywhere bounded  with probabiliy $1$, (as its equilibrium statistics is Gaussian, see part II),
the velocity $v(t)$ is a H\"older continuous function with exponent $H=1$ almost everwhere with probability 1.

\begin{figure}
\includegraphics[width=10cm]{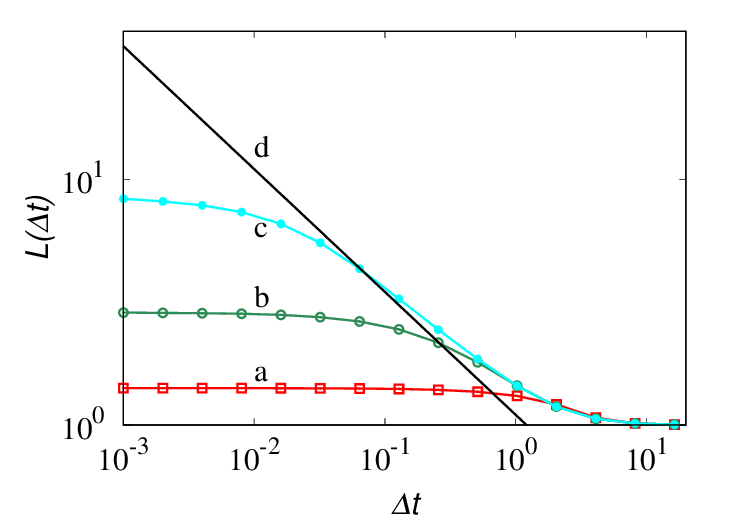}
\caption{Length-resolution analysis, $L(\Delta t)$ vs $\Delta t$, for a realization of
the ${\mathbb D}_{1,1}^{(1,1,0)}(\xi^{(d)}(t))$ patterns in the presence of a harmonic potential with $k=0.2$
at different values of $\lambda$.
Symbols  are the results of the stochastic simulations: ($\square$) and line (a): $\lambda=1$,
($\circ$) and line (b): $\lambda=10$, ($\bullet$) and line (c): $\lambda=100$. Line (d) represents
the Wiener scaling $L(\Delta t) \sim \Delta t^{-1/2}$.}
\label{Fig6}
\end{figure}

This is indeed a generic property of dissipative memory dynamics with finite relaxation rates $\lambda$,
in the absence of fluid-inertial effects. In other words, the presence of a linear viscoelastic
behaviour not only solves the paradox of the infinite speed of propagation
of the shear stresses (subsumed by the Newtonian constitutive equations), but also regularizes the
local structure  of the  velocity fluctuations of a colloidal
particle  immersed in the fluid.
Figure \ref{Fig6} depicts the  scaling of the length $L(\Delta t)$ vs $\Delta t$ for realizations
of this pattern at different values of $\lambda$. The value $\lambda=1$ corresponds to a relaxation time
equal to the dissipation time  $t_{\rm diss}$, and $\lambda=10\, ,100$ correspond to short relaxation time,  a tenth and a hundredth
smaller than  $t_{\rm diss}$, respectively. But even in this rather extreme cases, the saturation towards a constant value
of $L(\Delta t)$ is clearly evident starting from $\Delta t \sim 1/\lambda$ and reducing further $\Delta t$.

To conclude, consider the case of ${\mathbb D}{\mathbb I}_{1,1}^{(1,1,1)}(\xi^{(d)}(t),\xi^{(i)}(t))$, namely the coupling of fluid inertial effects
with a dissipative memory dynamics.  A single mode for each hydrodynamic mechanism is considered.
For this pattern (see part I), the dynamic equations read
\begin{eqnarray}
\dot{x}(t) & = & v(t) \nonumber \\
\dot{v}(t) & = & -\lambda \, \theta(t) - \gamma \, v(t) + \gamma \, \mu \, z(t) - k \,  x(t) + \sqrt{2\, \gamma} \,  \xi^{(i)}(t)
\nonumber \\
\dot{\theta}(t) & = & -\lambda \, \theta(t) + v(t) + \sqrt{2} \, \xi^{(d)}(t) \label{eqreg7} \\
\dot{z}(t) & = & - \mu \, z(t) + v(t) - \sqrt{\frac{2}{\gamma}} \, \xi^{(i)}(t) \nonumber
\end{eqnarray}
with $\xi^{(\beta)}(t)=d w^{(\beta)}(t)/dt$, $\beta=d,i$,  where  $w^{(d)}(t)$ and  $w^{(i)}(t)$ are Wiener processes independent of each other.

As can be observed from the structure of eqs. (\ref{eqreg7}), and even more clearly from
the block-diagram of the stochastic couplings  of this pattern depicted in part I, the presence of fluid inertia determines
a direct connection (proportionality) between the particle acceleration $\dot{v}(t)$ and $\xi^{(i)}(t)$. Consequently, 
$v(t) \sim w^{(i)}(t)$, from which it follows that the regularity properties of the particle velocity should be those
of a Wiener process. This implies that in ${\mathbb D}{\mathbb I}$-patterns we would expect a fractal velocity signal characterized by
a H\"older exponent $H=1/2$,  as in the case of the Stokesian ${\mathbb S}_{1,1}^{(1,1,0)}(\xi(t))$ pattern, indepedently
of the presence of a dissipative memory dynamics. This result is  generic as regards the number of modes considered, and it is
valid  {\em a fortiori}
 if we replace the dissipative memory dynamics with an instantaneous Stokesian friction.
This result is indeed a direct consequence of the
presence of an impulsive contribution in the
autocorrelation function of the stochastic force $R(t)$ associated
with the effects of fluid-inertial interactions, eqs. (63) and (64)
in part I. 

We can therefore conclude that, independently of the fluid-dynamic model adopted in determining
the fluid inertial contribution, stemming either from an incompressible model, or including the effects of compressibility,
as in the Chow and Hermans analysis, we would observe from the  time-series of particle velocities in ${\mathbb D}{\mathbb I}$-patterns
an almost everywhere singular  behaviour with a fractal dimension $d_v=3/2$. The  ${\mathbb D}{\mathbb I}$-pattern is the
most general  fluctuation-dissipation scheme deriving from the assumption of  the validity of the Langevin condition.
Its correlated generalization, namely the pattern 
${\mathbb C} {\mathbb D}{\mathbb I}_{m,E}^{(n,N_d,N_i)}({\xi}^{(d)}(t),\xi^{(i)}(t))$  removes this singularity
due to the presence of  a correlated stochastic forcing $q^{(i)}(t)$ in place of $\xi^{(i)}(t)$, acting
exactly in the same way as a viscoelastic memory term, and restores  a regular
behaviour of the particle velocity characterized by the H\"older exponent $H=1$.

\begin{figure}
\includegraphics[width=10cm]{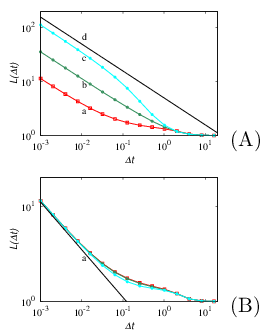}
\caption{Length-resolution analysis, $L(\Delta t)$ vs $\Delta t$, for a realization of
the ${\mathbb D} {\mathbb I}_{1,1}^{(1,1,1)}(\xi^{(d)}(t), \xi^{(i)}(t))$ patterns in the presence of a harmonic potential with $k=0.2$
and $\lambda=1$,
at different values of  the parameters characterizing the inertial term.
Symbols  are the results of the stochastic simulations.
Panel (A): $\mu=1$.  Symbols ($\square$) and line (a) refer  to $\gamma=0.1$, ($\circ$) and line (b) to $\gamma=1$,
($\bullet$) and line (c) to $\gamma=10$. The solid line (d)
represents the scaling $L(\Delta t) \sim \Delta t^{-1/2}$. Panel (B): $\gamma=0.1$.
 Symbols ($\square$) refer  to $\mu=0.1$, ($\circ$) to $\mu=1$,
($\bullet$) to $\mu=10$. The solid line (a)
represents the scaling $L(\Delta t) \sim \Delta t^{-1/2}$.}
\label{Fig7}
\end{figure}

Figure \ref{Fig7} panels (A) and (B) depict the fractal length-resolution analysis of velocity realizations arising from this
pattern. As predicted, a  singular scaling characterized by the fractal dimension $d_v=3/2$ of the Wiener 
processes is observed over the whole range of values of the inertial parameters $\mu$ and $\gamma$.
The effect of $\mu$ is practically negligible as can be observed from the data of panel (B), almost collapsing
into a single curve. The effect of $\gamma$ (panel A) is much more sensible and, as $\gamma$ decreases, the
resolution required to observe a fractal behaviour is below $\Delta t=0.1$, i.e. order of a tenth of the
dissipation time.

Therefore, the possible occurrence of regular velocity dynamics in Brownian motion
experiments, (this has been reported in \cite{exp2,deemed}, albeit it  should be further checked experimentally in order 
to obtain a definitive validation of it),
 strongly supports the thesis developed in this article of correlated forcings, at least as regards the inertial
terms $\boldsymbol{\xi}^{(i)}(t)$. But, as discussed in the previous Sections, this implies also the occurrence of a non vanishing
added mass stemming from correlated  stochastic forcings.

Back to the hydrodynamic reasoning, this result is not surprising if not intuitive: the
inclusion in particle hydromechanics of the effects of a finite speed of sound, i.e. compressibility,
represents solely a partial and incomplete refinement of the theory, as it keeps in classical fluctuational pattern theory, an infinite
propagation velocity of the stochastic forcings arising  a consequence  of the hydrodynamic fluctuations
in the fluid medium.  The complete regularization of the hydrodynamic interactions, removing
also the singularity of Wiener forcings  in fluctuation-dissipation models, not only ensures the regularity of $v(t)$,
but, as an aftermath, determines the emergence of the added-mass effect  in a way that is consistent with the global fluctuation-dissipation
relation.

\section{Fluctuation-inertia relation}
\label{sec6}

In the previous Sections we have outlined a fluctuational theory of the added-mass effect consistent with the
global fluctuation-dissipation relation and associated with correlated stochastic forcings acting within the
structure of the fluctuational patterns.

But the phenomenon of added mass is also a hydrodynamic property occurring in the accelerated motion of
macroscopic objects in a fluid medium of non-negligible density with respect to their densities.
It is therefore important to address the connection between the fluctuational added mass and its  macroscopic 
hydrodynamic  counterpart. For simplicty, consider the case of a spherical particle, amenable to
a scalar formalism.

The starting point is the structure of the force exerted  by the fluid on the particle
in a Newtonian incompressible fluid eq. (\ref{eq2_1}).  Apart from the specific instantaneous nature of
the dissipative kernel and of the component of the fluid-inertial kernel
pertaining to the instantaneous added mass, the  dependence of the hydromecanic kernels on
the physical parameters characterizing the fluid (density $\rho_0$ and viscosity $\mu$) reveals:
\begin{itemize}
\item the presence of a dissipative kernel $h(t)$ proportional to the fluid viscosity  $\mu$,
\begin{equation}
h(t) \sim \mu
\label{eqhh1}
\end{equation}
and consequenly vanishing in the limit $\mu \rightarrow \infty$ of an inviscid  fluid;
\item  the decomposition of the fluid inertial kernel into two contributions,
\begin{equation}
k(t) = k_\mu(t) + k_0(t)
\label{eqhh2}
\end{equation}
where $k_\mu(t)$ depends both on $\rho_0$ and $\mu$
\begin{equation}
k_\mu(t) \sim \sqrt{\mu \, \rho_0}
\label{eqhh3}
\end{equation}
vanishing for $\mu \rightarrow \infty$ and responsible, in eq. (\ref{eq2_1}),  for the
Basset force, and $k_0(t)$ proportional exclusively to the fluid density,
\begin{equation}
k_0(t) \sim \rho_0
\label{eqhh4}
\end{equation}
that in the Newtonian incompressible case accounts for the instantaneous added mass.
\end{itemize}
This classification can be extended to generic fluids, substituting the Newtonian viscosity
with the long-term viscosity as the integral of the viscosity memory kernel \cite{macosko}.
By the linearity of the hydrodynamic regimes considered, the decomposition eq. (\ref{eqhh2})  depends on the splitting
of the overall shear-stress tensor into dissipative shear stresses associated with $k_\mu(t)$,
and compressible pressure forces determining $k_0(t)$.
The latter contribution can be thus uniquely defined  starting from the overall
fluid inertial kernel $k(t)$ as
\begin{equation}
k_0(t)= \lim_{\mu \rightarrow 0} k(t)
\label{eqhh5}
\end{equation}
In the most accurate representation of fluid-particle interactions in which acoustic effects are also accounted for, $k_0(t)$ is  a continuous
function of its argument,  rapidly falling to zero for timescales comparable
with the characteristic dissipation time $t_{\rm diss} = m/(6 \, \pi  \, \mu \, R_p)$
(order of $10^{-6}$-$10^{-7}$ s, for a micrometric particle in water),
as the acoustic time scale $t_{\rm acou} \simeq R_p/c_s$,  is order of $10^{-9}$  s in
the same particle-fluid system.
Introducting the decomposition  eq. (\ref{eqhh2}) into particle
hydromechanics,
we have
\begin{eqnarray}
m \, \frac{d v(t)}{d t} & = & - \int_0^t h(t-\tau) \, v(\tau) \, d\tau - \int_0^t k_\mu(t-\tau) \, \left
( \frac{d v(\tau)}{d \tau}+ v(0) \delta (\tau) \right )  d \tau \nonumber \\
& + &  \int_0^t k_0(t-\tau) \, \left
( \frac{d v(\tau)}{d \tau}+ v(0) \delta (\tau) \right )  d \tau
\label{eqhh6}
\end{eqnarray}
Next, consider a macroscopic object slowly  accelerating in a fluid.
For the time-scales at which $k_0(t)$ is significantly different from zero, $d v(t)/dt$ is
practically a constant acceleration, and consequently the last term in eq. (\ref{eqhh6}), upon
a change of variables in the convolution integral, can be expressed as
\begin{equation}
\int_0^t k_0(\tau) \, \left
( \frac{d v(t- \tau)}{d \tau}+ v(0) \delta (t-\tau) \right )  d \tau
 \simeq \frac{d v(t)}{d t} \, \int_0^t k_0(\tau) \, d \tau \ \simeq \frac{d v(t)}{d t} \, m_a^{({\rm hydro})}
\label{eqhh7}
\end{equation}
where the hydrodynamic added mass $m_a^{({\rm hydro})}$ equals the integral of the inviscid kernel $k_0(t)$,
\begin{equation}
m_a^{({\rm hydro})} = \int_0^\infty k_0(t) \, d t
\label{eqhh8}
\end{equation}
From the above analysis it may seem that  $m_a^{({\rm hydro})}$ and the fluctuational
added mass entering the statistical properties of velocity fluctuations of micrometric particles eq. (\ref{eq3})
refer to two different hydrodynamic phenomenologies.
But the origin of the added-mass effect, either at microscale or for macroscopic objects, should stem
from one and the same physical mechanism. This is further   confirmed qualitatively (yet not fully quantitatively, 
thus a further careful experimental scrutiny is required)
by the experimental results reported in
\cite{optics_express}.

This observation leads to a principle of equivalence of the added masses, that can be referred to
as the {\em fluctuation-inertia} relation.  This is stated by the following  conjecture:\\

\vspace{0.2cm}
\noindent
{\bf Conjecture - Fluctuation-inertia relation} - The fluctuational added mass $m_a$, defined by eq. (\ref{eq3})
and the hydrodynamic added mass defined by eq. (\ref{eqhh8}), for the same object in the same
hydrodynamic conditions,  coincide. \\

\vspace{0.2cm}

Given the validity of the above principle, it remains to determine the structure of the  correlated stochastic forcings.
If we assume that each $q_i(t)$ is characterized by the same relaxation rate $\nu$ then, for any specific fluctuational
pattern, the  fluctuational added mass $m_a$ is a monotonically decreasing functon of $\nu$,
$m_a=\Phi(\nu)$ such that $\lim_{\nu \rightarrow \infty} \Phi(\nu)=0$.
Making use of the fluctuation-inertia principle, $\nu$ is uniquely defined by the relation
\begin{equation}
\nu= \Phi^{-1}\left ( m_a^{({\rm hydro})} \right )
\label{eqxxx}
\end{equation}
Of course, it is important to stress that the velocity autocorrelation function determined by
the correlated patterns enforcing eq. (\ref{eqxxx}) is different from the Kubo FD1k relation
based on the validity of the Langevin condition. However, 
once the theory is applied to realistic Prony expansions of hydromechanic kernels, it may happen
that the difference in the two autocorrelation functions may be tiny and difficult to detect from
experimental measurements.

Eq. (\ref{eqxxx}) represents a first closure of a correlated theory of  hydromechanics velocity fluctuations.
Further theoretical work is needed in order to improve this formulation in the
form of  fluctuation-inertia relations analogous to the Kubo first and second   fluctuation-dissipation
relations,  defining the fine structure of the correlated fluctuations and relating them to
functional structure of $k_0(t)$, and not only to its integral.

The fluctuation-inertia relation unifies the inertial phenomenology at micro- and macro\-scales. It physically indicates
that the characteristic correlation time associated with fluid fluctuations is related to the microscopic added mass, and vice versa.
For this reason, it can be regarded as the inertial counterpart of the global fluctuation-dissipation relation eq. (\ref{eq3_1}).
Further experiments on micrometric particles in liquids should be performed in order to verify this conjecture. On the other hand, the
proposed conjecture stresses also the  theoretical hydrodynamic modelling of a compressible
liquid phase as, the compressible approach proposed by Chow and Hermans  should be extended to a more
realistic description of a compressible liquid phase. To this purpose, the theory proposed in \cite{giona_ns} is
an interesting starting point.

\section{Concluding remarks}
\label{sec7}
This  third, and final article of the trilogy on fluctuation-dissipation
relations has addressed a new theoretical formulation of particle
hydromechanics grounded on the overcoming of the Langevin
condition,  
within the constraint imposed by the global
fluctuation-dissipation relation. Technically, this
is made possible by considering correlated stochastic forcing within
the basic fluctuational patterns introduced in part I.
This generalization follows the research line  
towards a progressively more accurate description of
the hydrodynamics  at microscale, removing the
unphysical approximations, such e.g. the assumption of incompressibility,
thus accounting for the finite speed of propagation of any hydrodynamic field
(pressure/density, internal stresses).

While the  inclusion of acoustic effects, still imbedded within a Wiener-based
theory of fluctuations, such e.g. in the Chow-Hermans work, has ``eliminated ''
the occurrence of the added-mass effect, restoring the  
energy equipartition for the squared velocity variance  of
colloidal particles in a still liquid, in  virtue of the regularity (finite speed of sound)
of the acoustic propagation,  the consistent, and almost due extension of the
same regularity principle, applied to stochastic forcings entering the equations
of motion, restored the added-mass effect as a generic property of correlated elementary fluctuations.
This restoration of the added-mass is altogeher different from the  same hydrodynamic
counterpart occuring due to incompressibility, as it occurs in generic compressible flows
due to the correlated properties of thermal/hydrodynamic fluctuations.

In the development of  the theory, the role of the fluctuational patterns introduced in part I
should be stressed, as the basic structural form for describing in a compact way FD3k in 
different hydrodynamic conditions of increasing  complexity.   
The structure of these patterns has been exploited in part II in order to analyze the properties
of white-noise impulsive stochastic forcings, and in the present article to extend the theory to
correlated fluctuations  maintaining the validity of the global fluctuation-dissipation
relation.

Moreover,  we have linked the occurrence of the added-mass effect, associated with
fluctuational properties, to the regularity of the velocity signal $v(t)$.
And also this property clearly emerges from the structure of the fluctuational
patterns accounting for fluid-inertial effects, i.e. ${\mathbb D}{\mathbb I}_{m,E}^{(n,N_d,N_i)}$.
While velocity signals in ${\mathbb D}{\mathbb I}_{m,E}^{(n,N_d,N_i)}$ displays a fractal
character (almost everywhere not differentiable), this property disappear in 
their  correlated counterparts, i.e. ${\mathbb C}{\mathbb D}{\mathbb I}_{m,E}^{(n,N_d,N_i)}$,
possessing Lipshitzian velocity signals.

The regularity of velocity signals (i.e. their almost everywhere Lipshitz continuity)
can be ``easily'' checked experimentally starting from
the analysis of velocity time series in highly resolved Brownian experiments.
It constitutes the local (in time) counterpart on the cumulative statistical
effect expressed by a non-vanishing fluctuational added mass.

A major thoretical effort is still to be performed in order to derive a finer
characterization of fluctuation-inertia relations conceptually similar to the FD1k and FD2k relations
of the classical dissipative Kubo theory grounded on the Langevin condition.

A general fluctuation-inertia principle has also been outlined, connecting the
fluctuational added mass occurring at microscales to the hydrodynamic added mass characterizing
the inertial properties of macroscopic objects accelerating in a liquid.

The proposed theory  is potentially rich of new theoretical implications as regards either statistical physics
or hydrodynamics of liquid phases.  These implications will be thoroughly addressed in a forthcoming communication.\\

\vspace{0.2cm}

\noindent
{\bf Acknowledgment - } This research  received financial support from ICSC---Centro Nazionale di Ricerca in High Performance Computing, Big Data and Quantum Computing, funded by European Union---NextGenerationEU. One of the authors (M.G.) is grateful to
M. G.  Raizen for precious discussions.

\appendix
\section{Moment analysis of the ${\mathbb C}{\mathbb I}_{m,E}^{(1,1,1)}$ pattern}

Consider the  second-order moments
associated with the dynamics variables  of the  correlated pattern ${\mathbb C} 
{\mathbb I}_{m,E}^{(1,1,1)}(q(t),q_1(t))$ analyzed in
paragraph \ref{subsec4_3}. 
Set
\begin{equation}
{\bf X}= (m_{vv}^*, m_{vz}^*, m_{vq}^*, m_{vq_1}^*, m_{zz}^*, m_{z1}^*, m_{zq1}^*)^T
\label{eqa1}
\end{equation}
The vector ${\bf X}$ of the steady-state (equilibrium) values is the solution of the linear system
\begin{equation}
{\bf A} \, {\bf X}=  {\bf b}
\label{eqa2}
\end{equation}
where
\begin{equation}
{\bf A}=
\left ( 
\begin{array}{ccccccc}
-(\gamma+\alpha) & \gamma & \sqrt{2} \, a & \sqrt{2} \,d & 0 & 0 &  0 \\
0 & \mu & 0 & 0 & -\mu & 0 & \sqrt{2} \, c \\
\mu & - (\gamma+\alpha + \mu) & 0 & \sqrt{2} \, c & \gamma & \sqrt{2} \, a & \sqrt{2} \, d \\
0 & 0 & -(\gamma+\alpha+\nu) & 0 & 0 &\gamma &0 \\
0 & 0 & 0 & -(\gamma+\alpha+\nu_1) & 0 & 0 & \gamma \\
0 & 0 & \mu & 0 & 0 & -(\mu+\nu) & 0 \\
0 & 0& 0& \mu & 0 & 0 & -(\mu+\nu_1)
\end{array}
\right )
\label{eqa3}
\end{equation}
and
\begin{equation}
{\bf b}= \left ( 0, 0, 0, - \frac{a \, \nu}{\sqrt{2}} , - \frac{d \, \nu_1}{\sqrt{2}}, 0, - \frac{c \, \nu_1}{\sqrt{2}} \right )^T
\label{eqa4}
\end{equation}
Solving this linear system the  expression for $m_{vv}^*=\langle v^2 \rangle_{eq}$
takes the form 
\begin{equation}
\frac{m \, \langle v^2 \rangle_{\rm eq}}{k_B \, T}
= \frac{N_0 + N_1 \, \alpha + N_2 \alpha^2}{D_0 + D_1 \, \alpha
+ D_2 \, \alpha^2 + D_3 \, \alpha^3}
\label{eqa5}
\end{equation}
\begin{eqnarray}
N_0 & = & \nu \, \nu_1 \, (\gamma+ \mu) \, (\mu+\nu_1) \, (\gamma+\mu+\nu)
\nonumber \\
N_1 & = & \nu \, (\mu+\nu) \, (\mu+\nu_1)^2+ \gamma
\left [ \mu^2 \, \nu+ \nu_1^2 \, (\mu+\nu) + \nu \, \nu_1 \, (2 \,\mu+\nu) \right ]
\nonumber  \\
N_2 & =  & \nu \, (\mu+\nu) \, (\mu+\nu_1) \nonumber \\
D_0 & = & \nu \, \nu_1  \, (\gamma+\mu) \, (\gamma+\mu+\nu) \, (\gamma + \mu+ \nu_1) 
\label{eqa6} \\
D_1 & = & \nu_1 \, (\gamma+ \mu) \, (\mu+\nu) \, (\gamma+\mu+\nu_1) +
\nu \, (\gamma+\mu) (\mu + \nu_1) \, (\gamma+ \mu+ \nu)
\nonumber \\
&+& \nu \, \nu_1 \,  (\gamma+ \mu+\nu) \, (\gamma+\mu+\nu_1) \nonumber \\
D_2 & = & (\gamma+\mu) \, (\mu+ \nu)\,(\mu+\nu_1) + \nu_1 \, (\mu + \nu)
\, (\gamma+\mu+ \nu_1) + \nu \, (\mu+\nu_1) \, (\gamma+\mu+\nu) \nonumber \\
D_3 & = & (\mu+\nu) \, (\mu+\nu_1) \nonumber
\end{eqnarray}
Set ${\bf p}=(\gamma,\mu,\nu,\nu_1)$ and let ${\bf p}^*=(\gamma^*,\mu^*,\nu^*,\nu_1^*)$ be the corresponding nondimensional set of  rates rescaled by $1/\alpha$,
i.e. $\gamma^*=\gamma/\alpha$ and similarly for the other entries.
The squared velocity variance at equilibrium can be expressed
as
\begin{equation}
\left . \frac{m \, \langle v^2 \rangle_{\rm eq}}{k_B \, T} 
= \frac{N_0 + N_1  + N_2}{D_0+D_1+D_2+D_3} \right |_{{\bf p}={\bf p}^*}
\label{eqa7}
\end{equation}

\end{document}